\documentstyle[floats,prd,aps,epsfig,eqsecnum,12pt]{revtex}
\makeatletter
\newbox\tempboxa
\newdimen\captionboxsubcount 
\def\capsize#1{\captionboxsubcount=#1pt}
\newdimen\captionboxsub
\captionboxsub=\hsize \advance\captionboxsub by -\captionboxsubcount
\advance\captionboxsub by -\captionboxsubcount
\long\def\@makecaption#1#2{
 \setbox\@tempboxa\hbox{#1 #2}
 \ifdim \wd\@tempboxa >\captionboxsub 
\rightskip=\captionboxsubcount \leftskip=\captionboxsubcount #1 #2 
\else \hbox to\hsize{\hfil\box\@tempboxa\hfil} 
 \fi}
\makeatother
\capsize{30}

\begin{document}
\bibliographystyle{unsrt}
\begin{titlepage}
\begin{flushright}
\begin{minipage}{5cm}
\begin{flushleft}
\small
\baselineskip = 13pt
hep-ph/0012278 \\
\end{flushleft}
\end{minipage}
\end{flushright}
\vskip 4.0cm
\begin{center}
\Large\bf
Unitarized pseudoscalar meson scattering amplitudes from three flavor
linear sigma models.
\end{center}
\footnotesep = 12pt
\vskip 2.0cm
\begin{center}
\large
Deirdre {\sc Black}$^{\it \bf a}$ 
\footnote{Electronic address: {\tt black@physics.syr.edu}}
\quad\quad Amir H. {\sc Fariborz}$^{\it \bf {b}}$ \footnote{Electronic  
address: {\tt fariboa@sunyit.edu}}\\
\vskip 0.5cm
Sherif {\sc Moussa}$^{\it \bf a}$ \footnote{Electronic address:{\tt sherif@suhep.phy.syr.edu}} \quad\quad 
Salah {\sc Nasri}$^{\it \bf a}$ \footnote{Electronic address: {\tt
snasri@suhep.phy.syr.edu}}\\
\vskip 0.5cm
Joseph {\sc Schechter}$^{\it \bf a}$ \footnote{Electronic address : {\tt schechte@suhep.phy.syr.edu}}\\
{\it 
\qquad $^{\it \bf a}$ Department of Physics, Syracuse University, 
Syracuse, NY 13244-1130, USA.} \\
{\it 
\qquad $^{\it \bf b}$ Department of Mathematics/Science, State University of New York Institute of Technology, 
Utica, NY 13504-3050, USA.} \\
\end{center}
\vskip 8.0cm
\begin{center}
\bf
Abstract
\end{center}
\begin{abstract}
The three flavor linear sigma model is studied as a ``toy model'' for understanding the role of possible light scalar mesons in the $\pi \pi$, $\pi K$ and $\pi \eta$ scattering channels.  The approach involves computing the tree level partial wave amplitude for each channel and unitarizing by a simple K-matrix prescription which does not introduce any new parameters.  If the renormalizable version of the model is used there is only one free parameter.  While this highly constrained version has the right general structure to explain $\pi \pi$ scatteirng, it is ``not quite'' right.  A reasonable fit can be made if the renormalizability (for the {\it effective} Lagrangian) is relaxed while chiral symmetry is maintained.  The occurence of a Ramsauer Townsend mechanism for the $f_0(980)$ region naturally emerges.  The effect of unitarization is very important and leads to ``physical'' masses for the scalar nonet all less than about 1 GeV.  The $a_0(1450)$ and $K_0^*(1430)$ appear to be ``outsiders'' in this picture and to require additional fields.  Comparison is made with a scattering treatment using a more general non-linear sigma model approach.  In addition some speculative remarks and a highly simplified larger toy model are devoted to the question of the quark substructure of the light scalar mesons.  
\baselineskip = 17pt
\end{abstract}
\begin{flushleft}
\footnotesize
PACS number(s): 13.75.Lb, 11.15.Pg, 11.80.Et, 12.39.Fe 
\end{flushleft}
\vfill
\end{titlepage}
\setcounter{footnote}{0}
\section{Introduction}
In the last few years there has been a revival of interest \cite{vanBev}-\cite{proc} in the
possiblity that light scalar mesons like the sigma and kappa exist.
This is a very important but highly controversial subject.  The
difficulty is that one must demonstrate their existence by comparing
with experiment, believable theoretical amplitudes containing the
light scalars.  However, the energy range of interest is too low for
the systematic perturbative QCD expansion and too high for the
systematic chiral perturbation theory expansion \cite{CPT}.  Clearly, chiral
symmetry should hold but it seems unavoidable to fall back on model
dependent approaches.  Qualitatively, the dominance of tree amplitudes is suggested by the $\frac{1}{N_c}$ expansion \cite{LargeN} and it
has been shown by the Syracuse group \cite{SS,HSS1,BFSS1,BFS2} that this approach can be used to
economically fit the data in the framework of a non-linear chiral
Lagrangian which includes vectors and scalars in addition to the
pseudoscalars.  Many related approaches have been discussed by other
workers \cite{Other}.  To put the problem in historical perspective, the theoretical
treatment of meson-meson scattering has been a topic of great interest
for about forty years and has given rise, among other things, to
chiral perturbation theory and string theory.  Nevertheless, the
problem itself of explaining light meson scattering amplitudes from
threshold to (say) about the 1.5 GeV region is still not definitively
solved.  Of course, if the existence of light scalars is true, it will
be a crucial step forward.  

In such a situation, it is often useful to increase one's perspective
by studying simplified ``toy models''.  The classic chiral symmetric
model which contains a scalar meson is the Gell-Mann L\'evy two flavor
linear sigma model \cite{GL}.  At tree level it yields essentially the same $\pi \pi$
scattering length which is the initial approximation in the chiral
perturbation scheme.  However, compared to that scheme, which uses a
non-linear Lagrangian of pions only \cite{Weinberg}, it is less convenient to
systematically implement corrections.  Nevertheless it does contain a
light scalar meson and it does provide the standard intuitive picture
of spontaneous chiral symmetry breaking.  Furthermore, it is likely to
be \cite{Hatsuda} an exact model close to the QCD chiral phase
transition.  Of course there is an enormous literature on the
application of the two flavor linear sigma model to $\pi \pi$
scattering.  Recently, Achasov and Shestakov \cite{AS} have shown that a
qualitatively reasonable picture emerges at the lower part of our
energy range of interest \cite{CH} by using a scheme which is
equivalent to what we may call ``K-matrix unitarization''.  Namely, in
the standard parameterization \cite{Chung} of a given partial wave S-matrix:
\begin{equation}
S = \frac {1 + iK}{1 - iK} \equiv 1 + 2iT,
\label{regularization}
\end{equation}
we identify 
\begin{equation}
K=T_{\rm{tree}}.
\label{Kdef}
\end{equation}
$T_{\rm{tree}}$ is the given partial wave T-matrix computed at tree
level and is purely real.  Such a scheme gives exact unitarity for T
but violates the crossing symmetry which $T_{\rm{tree}}$ itself obeys. 

For a more realistic application to $\pi \pi$ scattering
(i.e. inclusion of the $f_0(980)$) as well as to $\pi K$, $\pi\eta$
scatterings etc. it is highly desirable to extend this calculation to
the three flavor case.  That is the purpose of this paper.  We will see that it provides a very predictive and reasonably successful model which gives interesting new insights.

The three flavor linear sigma model \cite{Levy} is constructed from the $3 \times
3$ matrix field
\begin{equation}
M=S+i\phi
\end{equation}
where $S=S^\dagger$ represents a scalar nonet and $\phi= \phi^\dagger$
a pseudoscalar nonet.  Under a chiral transformation $q_L \rightarrow
U_L q_L$, $q_R \rightarrow U_R q_R$ of the fundamental left and right
handed light quark fields, M is defined to transform as 
\begin{equation}  
M \longrightarrow U_L M U_R^\dagger.
\end{equation}
To start with, one may consider a general non-renormalizable \cite{SU1}
Lagrangian of the form
\begin{equation}
{\cal L} = - \frac{1}{2} {\rm Tr} \left( \partial_\mu \phi \partial_\mu \phi
\right) - \frac{1}{2} {\rm Tr} \left( \partial_\mu S \partial_\mu S 
\right) - V_0 - V_{SB},
\label{LsMLag}
\end{equation}
where $V_0$ is an arbitrary function of the independent $SU(3)_L
\times SU(3)_R \times U(1)_V$ invariants
\begin{equation}
I_1 = {\rm Tr} \left( M M^\dagger \right), \quad  I_2 = {\rm Tr} \left( M
M^\dagger M M^\dagger \right), \quad I_3 = {\rm Tr} \left( (M
M^\dagger)^3 \right), \quad I_4 = 6\left( {\rm det} M + {\rm det}
M^\dagger \right).
\end{equation}
Of these, only $I_4$ is not invariant under $U(1)_A$.  The symmetry
breaker $V_{SB}$ has the minimal form 
\begin{equation}
V_{SB} = -2 \left( A_1 S_1^1 + A_2 S_2^2 + A_3 S_3^3 \right),
\end{equation}
where the $A_a$ are real numbers which turn out to be proportional to
the three light (``current'' type) quark masses.  

The Lagrangian Eq. (\ref{LsMLag}) contains the most general ``potential''
term $V_0$  but still has the minimal ``kinetic'' term.  It is
possible \footnote{See for example, section IV of \cite{GJJS} }
to also include non-renormalizable kinetic-type terms like ${\rm Tr}
\left( \partial_\mu M \partial_\mu M^\dagger M M^\dagger \right)$, 
${\rm Tr} \left( \partial_\mu M  M^\dagger \partial_\mu M M^\dagger
\right) + h.c.$, etc.  We shall disregard such terms in the present
paper.  It is interesting to note \cite{SU1} that the results of
``current algebra'' can be derived from Eq. (\ref{LsMLag}) without
knowing details of $V_0$, just from chiral symmetry and the assumption
that the minimum of $V \equiv V_0 + V_{SB}$ is non-zero;  specifically
the ``vacuum values'' satisfy 
\begin{equation}
\left< S_a^b \right> = \alpha_a \delta_a^b.
\end{equation}

The ``one-point'' vertices (pseudoscalar decay constants) are related
to these parameters by 
\begin{equation}
F_\pi = \alpha_1 + \alpha_2, \quad F_K = \alpha_1 + \alpha_3.
\label{decayconstants}
\end{equation}
In the isotopic spin invariant limit one has,
\begin{equation}
A_1 = A_2, \quad \alpha_1 = \alpha_2 \quad ({\rm isospin \,
limit}).
\end{equation}

Many, though not all, of the ``two-point'' vertices (particle squared
masses) may be calculated by \cite{SU1} single
differentiation of two ``generating functions'' which express the
chiral symmetry of $V_0$ and also using
\begin{equation}
\left< \frac {\partial V} {\partial S_a^b} \right> = 0.
\label{extremum}
\end{equation}
For example, one finds
\begin{equation}
m^2 \left( \pi^+ \right) = 2 \frac {A_1 + A_2}{\alpha_1 + \alpha_2},
\quad m^2 \left( K^+ \right) = 2 \frac {A_1 + A_3}{\alpha_1 +
\alpha_3}.
\label{piKmasses}
\end{equation}
The formula for the mass of the $\eta^\prime$ (and of the particles
$\eta$ and $\pi^0$ with which it may mix) also involves the quantity 
\begin{equation}
V_4 \equiv \left< \frac{ \partial V_0} {\partial I_4} \right>.
\end{equation}

Many of the three point and four point vertices may be obtained by
respectively two times and three times differentiating the above
mentioned generating equations.  The specific terms needed for our
subsequent discussion are given in Appendix A.  

The present model requires us (in the limit of isospin invariance) to
specify the five parameters, $A_1$, $A_3$, $\alpha_1$, $\alpha_3$ and
$V_4$.  These may be obtained by using the five experimental input
values:
\begin{eqnarray}
m_\pi = 0.137 {\rm GeV} &,& \, m_K = 0.495 {\rm GeV},  \nonumber \\
m_\eta = 0.457 {\rm GeV}&,& \, m_{\eta^\prime} = 0.958 {\rm GeV},
\nonumber \\
F_\pi & = & 0.131 {\rm GeV}  \quad \quad \left({\rm Inputs}\right) .  
\label{inputs}
\end{eqnarray}
This is a reasonable, but clearly not unique choice for the inputs.
(For example, $F_K$ might be used instead of $m_{\eta}$).  With these input parameters there are two immediate predictions for
pseudoscalar properties:
\begin{equation}
\theta_p = 2.05^o, \quad \frac{F_K}{F_\pi} = 1.39,
\label{predictions1}
\end{equation}
where the pseudoscalar mixing angle, $\theta_p$, is here defined by 
\begin{equation}
\left( \begin{array}{c} \eta \\ \eta^\prime \end{array} \right) = \left(
\begin{array}{c c} {\rm cos} \theta_p & -{\rm sin} \theta_p \\ {\rm sin}
\theta_p & {\rm cos} \theta_p \end{array} \right) \left( \begin{array}{c}
\eta_8 \\ \eta_0 \end{array} \right).
\label{psmixing}
\end{equation}
$\eta$ and $\eta^\prime$ are the ``physical'' states while the
``unmixed'' states are $\eta_8 = (\phi_1^1 + \phi_2^2 - 2
\phi_3^3)/{\sqrt{6}}$ and  $\eta_0 = (\phi_1^1 + \phi_2^2 +
\phi_3^3)/{\sqrt{3}}$.  The predictions in Eq. (\ref{predictions1}) are
qualitatively reasonable but not very accurate;  the usually accepted
value for $\theta_p$, while small, is \cite{MS} around $-18^o$ and the
experimental value for $\frac{F_K}{F_\pi}$ is about 1.22.  (It is
likely that the inclusion of non-renormalizable kinetic terms
mentioned above will improve this aspect).  

Now, the scalar meson masses are of the most present interest.
Analogously to the pseudoscalars ($\pi$, $K$, $\eta_0$, $\eta_8$) we
denote the scalars by ($a_0$, $\kappa$, $\sigma_0$, $\sigma_8$).  The
predicted squared masses from the ``toy'' Lagrangian in
Eq. (\ref{LsMLag}) are:
\begin{eqnarray}
\left( \alpha_2 - \alpha_1 \right) m^2_{\rm BARE} (a_0^+) &=& 2 (A_2 -
A_1), \nonumber \\
m^2_{\rm BARE} (\kappa^+) &=& \frac{2 (A_3 - A_1)}{\alpha_3 -
\alpha_1},  \nonumber \\
m^2_{\rm BARE} (\kappa^0) &=& \frac{2 (A_3 - A_2)}{\alpha_3 -
\alpha_2}.
\label{scalarmasses}
\end{eqnarray}
In this case, the masses of $\sigma_0$, $\sigma_8$ and their mixing
angle $\theta_s$ [defined analogously to Eq. (\ref{psmixing})] are {\it {not}}
predicted.  In the isotopic spin invariant limit, which we shall adopt
here, the $a_0$ mass is not predicted (although it may be reasonably
estimated \cite{SU2} by taking isospin violation into account).  Note
that we have, in contrast to the pseudoscalar case, put a subscript
``BARE'' on each scalar mass.  This is because the pole positions in
the pseudoscalar-pseudoscalar scattering amplitudes corresponding to
scalar mesons may be non-trivially shifted by the unitarization
procedure of Eqs. (\ref{regularization}) and (\ref{Kdef}).  We
consider the unitarization to be an approximation to
including all higher order corrections.  Then, in the usual field
theoretic way of thinking, the pole position
determines the physical mass, while the tree level $m_{\rm BARE}$ has
no clear physical meaning. 

The tree level $\pi \pi$ scattering amplitude is easily computed
\cite{SU1} from Eq. (\ref{LsMLag}) in the present scheme.  It
involves a four point ``contact'' amplitude and $\sigma$ and
$\sigma^\prime$ exchange diagrams.  The resulting form \footnote{The
sign of $A(s,t,u)$ is the negative of the one in the convention of \cite{SU1} but
in agreement with those in \cite{HSS1,HSS2,BFS1}.} turns
out to be remarkably simple:
\begin{eqnarray}
A \left( s,t,u \right) &=& \frac{2}{{F_\pi}^2} \left\{ m_\pi^2 + {\rm cos}^2\psi
\left[ \frac { {\left( m_{\rm BARE}^2\left( \sigma \right) - m_\pi^2
\right)}^2}{m_{\rm BARE}^2 \left( \sigma \right) - s} - m_{\rm BARE}^2
\left( \sigma \right) \right] \right. \nonumber \\ 
 &+& {\rm sin}^2\psi
\left[ \frac { {\left( m_{\rm BARE}^2(\sigma^\prime) - m_\pi^2
\right)}^2}{m_{\rm BARE}^2(\sigma^\prime) - s} - m_{\rm BARE}^2(\sigma^\prime)
\right] \left. \right\},
\label{pipiampl}
\end{eqnarray}
where $s$, $t$ and $u$ are the usual Mandelstam variables.  The angle
$\psi$ is defined by the transformation:
\begin{equation}
\left( \begin{array}{c} \sigma\\ \sigma^\prime \end{array} \right) = \left(
\begin{array}{c c} {\rm cos} \psi & -{\rm sin} \psi \\ {\rm sin}
\psi & {\rm cos} \psi \end{array} \right) \left( \begin{array}{c}
\frac{S_1^1 + S_2^2}{\sqrt{2}} \\ S_3^3 \end{array} \right),
\label{scalarmixing}
\end{equation}
where \footnote{Note that neither $\psi$ nor $\theta_s$ are defined in
the same way as $\theta_s$ in Eq. (3.6) of \cite{BFSS2}} $\psi$ is
related to the angle $\theta_s$ [defined analogously to Eq. (\ref{psmixing})] by
\begin{equation}
{\rm cos} \psi = \frac{1}{\sqrt{3}} \left( {\rm cos} \theta_s -
\sqrt{2} {\rm sin} \theta_s \right),
\end{equation}
which translates to $\psi \approx \theta_s + 54.7^o$.
With the Lagrangian Eq. (\ref{LsMLag}) the amplitude
Eq. (\ref{pipiampl}) depends on the three unknown parameters $m_{\rm BARE}(\sigma)$, $m_{\rm BARE}(\sigma^\prime)$ and $\psi$.  

\begin{figure}
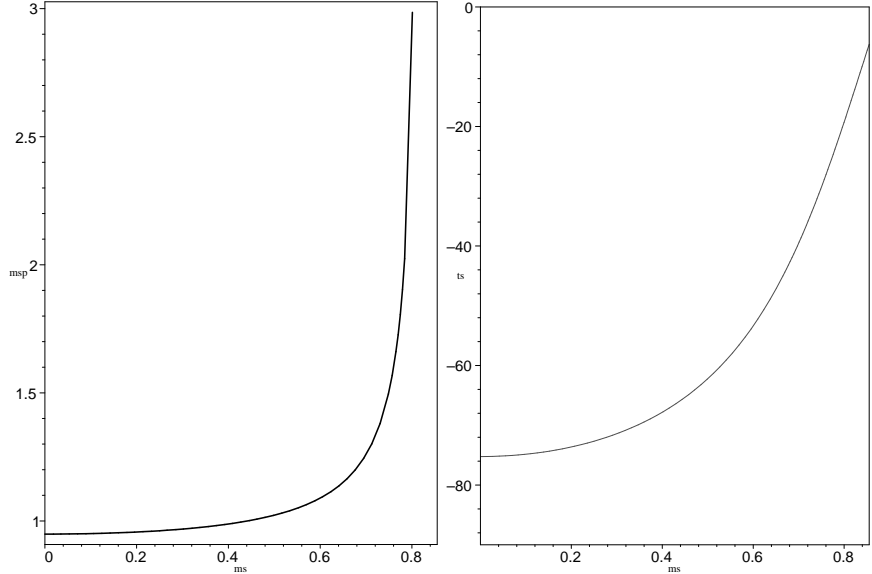

\centering
\epsfig{file=./ms_mf.eps, height=3in, angle=0}
\epsfig{file=./ms_thetas.eps, height=3in, angle=0}
\caption
{Dependence (left) of $m_{\rm \tiny BARE}(\sigma^\prime)$ in GeV and
(right) of the octet-singlet mixing angle $\theta_s$ in degrees on $m_{\rm
\tiny BARE}(\sigma)$, in the renormalizable linear sigma model.}
\label{ms_mf}
\end{figure}

We can increase the predictivity of the model by restricting the
potential $V_0$ in Eq. (\ref{LsMLag})  to contain only renormalizable
terms.  The resulting model is the one usually considered since it
allows for a consistent perturbation treatment (although the coupling
constants are very large).  In any
event, we will be working at tree level and ``simulating'' higher
order corrections by the K-matrix unitarization procedure.  Note that
all the formulas gotten above with general $V_0$ continue to hold in
the renormalizable model;  there will just be additional restrictions.
The renormalizable potential may be written \cite{HS} as:
\begin{equation}
V_0 ({\rm renormalizable}) = \left[ V_1 - V_{11} \left( \Sigma_a
\alpha_a^2 \right) \right] I_1 + \frac{1}{2} V_{11} {\left(I_1
\right)}^2 + V_2 I_2 + V_4 I_4,
\label{renorm_pot}
\end{equation}
where we have used the notation
\begin{equation}
V_a \equiv \left< \frac {\partial V_0}{\partial I_a} \right>, \quad
V_{ab} \equiv  \left< \frac {{\partial}^2 V_0}{\partial I_a \partial
I_b} \right>.
\end{equation}
As discussed \footnote{Please notice the relevant typographical
errors in \cite{HS}: 1) In the first of Eqs. (2.2), $A_3/w - 1$ should be replaced
by $A_3/w - A_1$.  2) In the second line of Eq. (2.5c) $4\alpha w
V_{11}$ should be replaced by $4{\alpha}^2 w V_{11}$.  3) In the
numerator of Eq. (2.8) the factor ${(4\alpha)}^{-2}$ should be
replaced by ${(2\alpha)}^{-2}$.} in \cite{HS}, we may determine
$V_1$ and $V_2$ from the extremum equation Eq. (\ref{extremum}) while
$V_{11}$ may be expressed in terms of $m_{\rm BARE}(\sigma)$.  Thus
specifying $m_{\rm BARE}(\sigma)$ determines the model parameters
completely.   Actually, $m_{\rm BARE}(a_0)$ is fixed to be 0.913 GeV
just by requiring renormalizability, independent of the choice of
$m_{\rm BARE}(\sigma)$.  Using Eq. (\ref{scalarmasses}) we find that
$m_{\rm BARE}(\kappa) = 0.909$ GeV, independent of whether we make the
renormalizability restriction or not.  Finally, the dependences of
$m_{\rm BARE}(\sigma^\prime)$ and $\theta_s$ on the choice of $m_{\rm
BARE}(\sigma)$ are displayed in Fig. \ref{ms_mf}.  Choosing the convention where $m_{\rm
BARE}(\sigma) < m_{\rm BARE}(\sigma^\prime)$, the model does not allow
for $m_{\rm BARE}(\sigma)$ greater than about 0.813 GeV.  Furthermore
$m_{\rm BARE}(\sigma^\prime)$ must be greater than about 0.949 GeV in
the renormalizable model.  

In section II we study the simpler two flavor model both for the purpose of review and for introducing our notation and the method we will use in the three flavor case.  We will also illustrate just how well the amplitude can be approximated by a pole in the complex s-plane plus a constant.  A number of new remarks are made.  Section III contains a detailed discussion of the s-wave $\pi \pi$, $\pi K$ and $\pi \eta$ scattering amplitudes in the unitarized three flavor model.  Both plots of the predicted amplitudes compared with experiment and numerical calculation of the pole parameters will be seen to be useful for understanding the dynamics.  A summary and discussion of the calculations of the scalar meson parameters are presented in section IV.  Section V contains a more speculative discussion on the question of the ``quark substructure'' of the light scalars.  It is pointed out that there is a difference in describing this at the ``current'' and ``constituent'' quark levels.  Also, while the linear sigma model is set up on the ``current'' quark basis, it does not uniquely describe the quark substructure.  In the present model, the initial ``current-quark'' meson field leads to constituent type states which are modified both by details of symmetry breaking and by unitarization.  The possible richness of the scalar meson system for further study is illustrated by the introduction of a larger toy model which includes two different $M$ matrices.  

\section{Two flavor linear sigma model}

It seems useful to first review the two flavor case and to make some
additional comments.  We start by exploring the difficulty with a
conventional extension of the tree level amplitude beyond the
threshold region.  This also provides the usual motivation for the
introduction of the non-linear sigma model.

\subsection{Standard unitarization procedure and its problems}

It is easy to get the two flavor $\pi \pi$ scattering amplitude by
taking a suitable limit of the three flavor amplitude given in
Eq. (\ref{pipiampl}).  We simply decouple the $\sigma^\prime$ by
setting the
$\sigma-\sigma^\prime$ mixing angle $\psi$ to zero, as is evident in Eq. (\ref{scalarmixing}).  Then $\sigma$
becomes $\frac{S_1^1 + S_2^2}{\sqrt 2}$, while $\sigma^\prime = S_3^3$
does not belong to the SU(2) theory and decouples; we are left with the tree
amplitude \footnote{Since this formula was gotten as a limit of the
SU(3) model with an arbitrary (not neccessarily a fourth order
polynomial) potential, we see that the tree result
Eq. (\ref{SU2pipiampl}) is independent of whether the
SU(2) linear sigma model potential has only renormalizable terms.}:
\begin{equation}
A \left( s,t,u \right) = \frac{2}{{F_\pi}^2} \left[ m_{\rm BARE}^2
\left( \sigma \right) - m_\pi^2 \right] \left[ \frac{m_{\rm BARE}^2
\left( \sigma \right) - m_\pi^2}{m_{\rm BARE}^2\left( \sigma \right) -
s} -1 \right].
\label{SU2pipiampl}
\end{equation}
The pole term in the second bracket represents the $\sigma$ exchange
Feynman diagram.  Naively one would expect this term by itself to
describe $\sigma$ dominance of the low energy amplitude.  However, the
$(-1)$ piece, which comes from the four point contact interaction,
is needed in this model to satisfy chiral symmetry.  It is easy to see
that there is a dramatic partial cancellation of the two terms near
threshold.  For example, if we take the single unknown parameter in
the model, $m_{\rm BARE} (\sigma)$ to be 1 GeV, then the pole term at
threshold $\left[ s_{th} = 4m_\pi^2 \right]$ is about 1.06.  This
gets reduced to just $6 \%$ of its value after adding the constant
piece.  Near threshold we can approximate $m_{\rm BARE}^2 (\sigma) \gg
\left[ m_\pi^2,s \right]$ to get the famous ``current algebra''
\cite{W} formula
\begin{equation}
A (s,t,u) \approx \frac { 2 \left( s - m_\pi^2 \right)}{F_\pi^2}.
\label{currentalgebra}
\end{equation}
We have just seen that this is a small quantity which has arisen from a
partial cancellation of two relatively large terms.  Now if we wish to
use Eq. (\ref{SU2pipiampl}) away from threshold we run into the
problem of an infinity arising when $s= m_{\rm BARE}^2 (\sigma)$.  A
standard unitarization procedure to avoid this problem would
correspond to making the replacement 
\begin{equation}
\frac{1} {m_{\rm BARE}^2 (\sigma) - s} \longrightarrow \frac{1}
{m_{\rm BARE}^2 (\sigma) - s - i\Gamma m_{\rm BARE} (\sigma) },
\label{conventionalreg}
\end{equation}
where $\Gamma$ is a width factor.  The trouble is that the delicate
partial cancellation with the contact term is now spoiled near
threshold and consequently there will be a very poor agreement with
experiment in the threshold region.  

The most popular alternative treatment introduces a non-linearly
transforming pion field and no $\sigma$ at all.  (Formally it may be
gotten by ``integrating out'' the $\sigma$ of the linear model but
this is not the most general formulation).  Then the current algebra
formula Eq. (\ref{currentalgebra}) is obtained directly from a
derivative type four point contact term (as opposed to the
non-derivative type in the linear model).  This approach forms the
basis of the chiral perturbation scheme (of pions only).  The next
order correction will involve more powers of derivatives and hence
will not drastically modify the already reasonable current algebra
result. 

A sigma-type particle can be introduced in a general way (independent
of the linear sigma model) in the non-linear framework by using a
standard technique \cite{CCWZ}.  In this approach the $\sigma \pi \pi$ couplings
are inevitably of derivative type so the $\sigma$-pole contribution is
small near threshold and does not drastically alter the current
algebra result.  This is clearly convenient since a regularization of
the type Eq. (\ref{conventionalreg})  will not now alter the threshold
behavior drastically.  However, this does not neccessarily guarantee
good experimental agreement away from threshold.  

It seems worthwhile to emphasize that both the linear and non-linear
sigma models represent the same physics - spontaneous breakdown of
chiral symmetry.  The choice of which to use is hence primarily a
question of convenience in extending the description away from
threshold.  In this paper we focus on studying the linear model,
regarding it as a ``toy model'' useful for increasing our understanding.  

To go further, we need the partial wave projection ot the amplitude
Eq. (\ref{SU2pipiampl}).  Here we specialize to the I=0 projection:  
\begin{equation}
A^{\rm{I}=0} = 3 A\left( {\rm{s,t,u}} \right) + A\left( {\rm{u,t,s}} \right)
+ A\left( {\rm{t,s,u}} \right).
\label{pipi-isospinprojection}
\end{equation}
The angular momentum $l$ partial wave elastic scattering amplitude for isospin $I$ is \
\begin{equation}
T_{l}^I (s) =  \frac{1}{2} \rho(s) \int^1_{-1} d{\rm cos} \theta \, P_l \left(
{\rm cos} \theta \right) \, A^I (s,t,u),
\label{partialwave}
\end{equation} 
where $A^I(s,t,u)$ is the isospin I invariant amplitude, $\theta$ is the center of mass scattering angle and 
\begin{equation}
\rho(s) = \frac {q(s)}{16\pi \sqrt s},
\label{kinematicalfactor}
\end{equation}
with $q(s)$ the center of mass momentum for, in general, a channel containing particles $a_1$ and $a_2$:
\begin{equation}
q^2 = \frac{s^2 + {\left( m_{a_1}^2 - m_{a_2}^2 \right)}^2 - 2s \left(
m_{a_1}^2 + m_{a_2}^2 \right)}{4s}.
\label{cofm}
\end{equation}
$T_l^I$ is related to the partial wave S-matrix, $S_I^l$ by
Eq. (\ref{regularization}).  For understanding the properties of the
$\sigma$-meson, the $T_0^0$ amplitude is clearly the most relevant.
Using Eq. (\ref{SU2pipiampl}) and
Eqs. (\ref{pipi-isospinprojection})-(\ref{cofm}) we get the tree
approximation
\begin{equation}
{T^{0}_{0}}_{tree}(s) =  \alpha \left( \rm{s}
\right) + \frac{\beta (s)}{{m_{\rm BARE}^2(\sigma)} - s}
\label{pipi-partialwave}
\end{equation}
where 
\begin{eqnarray}
\alpha \left( \rm{s}\right) &=& \rho (s) \frac{{m_{\rm BARE}^2 (\sigma)} - {m_\pi}^2}{{F_\pi}^2}
\left[ -10 + 4 \frac{{m_{\rm BARE}^2(\sigma)} - {m_\pi}^2}{s - 4 {m_\pi}^2} \rm{ln}
\left( \frac{{m_{\rm BARE}^2(\sigma)} + s - 4{m_\pi}^2}{{m_{\rm
BARE}^2(\sigma)}} \right) \right],  \nonumber \\
\beta (s) &=& \frac{6 \rho (s)}{{F_\pi}^2} {\left( {m_{\rm BARE}^2(\sigma)} - {m_\pi}^2 \right)}^2.
\label{alphabeta}
\end{eqnarray}
Note that $\alpha (s)$ in Eq. (\ref{alphabeta}) does {\it not} blow up
when $q^2 = \frac {s- 4m_\pi^2}{4} \rightarrow 0$.

\begin{figure}
\centering
\epsfig{file=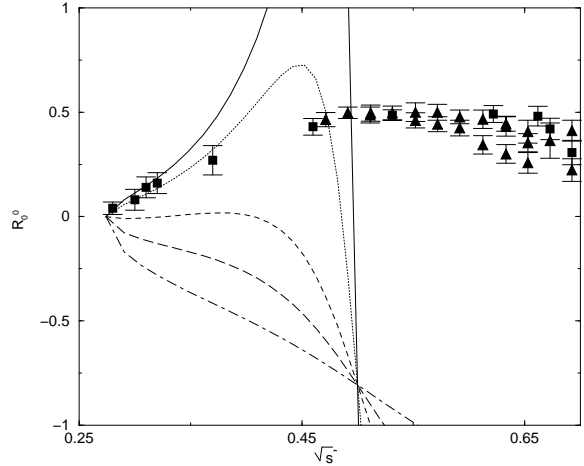, height=3in, angle=270}
\caption
{Predicted Real part of $\pi \pi$ I=0 s-wave amplitude using
regularization with different constant widths according to
Eq. (\ref{conventionalreg}).  The widths are 50 MeV (solid), 100 MeV (dots), 200 MeV (dashes), 300 MeV
(long-dashes), 500 MeV (dot-dashes). Here, $m_{\rm \tiny BARE}
(\sigma)= 560$ MeV.}
\label{SU2LsMconstwidths_fig}
\end{figure}

\begin{figure}
\centering
\epsfig{file=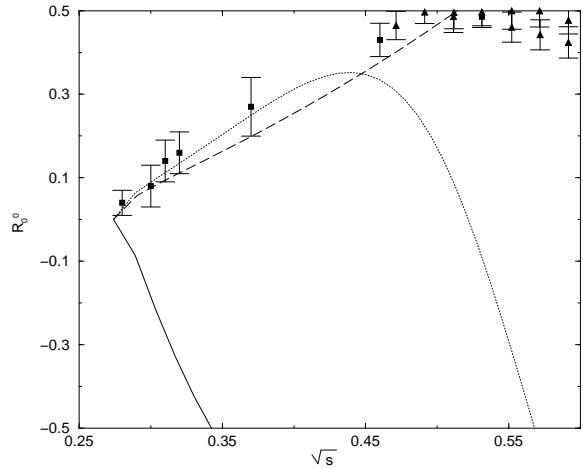, height=3in, angle=270}
\caption
{Variations on SU(2) linear sigma model prediction for Real part of
$\pi \pi$ I=0 s-wave amplitude.  The dashed curve is the ``current
algebra'' result Eq. (\ref{currentalgebra}), the solid curve uses
Eq. (\ref{conventionalreg}) with the constant perturbative width
calculated from Eq. (\ref{sigmawidth}) and the dotted curve uses the
regularization prescription outlined around Eq. (\ref{Amirwidth}).
Here, $m_{\rm \tiny BARE}(\sigma) = 560$ GeV.}
\label{SU2LsMwidthsrun_fig}
\end{figure}

\begin{figure}
\centering
\epsfig{file=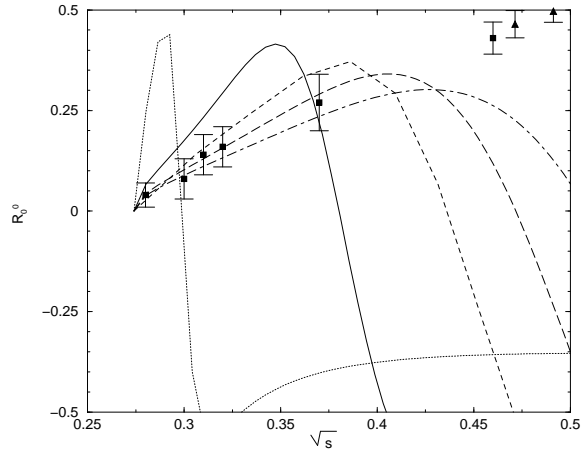, height=3in, angle=270}
\caption
{Effect of different values of $m_{\rm \tiny BARE}(\sigma)$ on the
generalized ``running width'' prescription, outlined around
Eq. (\ref{Amirwidth}), on the SU(2) linear sigma model prediction for
Real part of the $\pi \pi$ I=0 s-wave scattering amplitude.  We show
the curves for (in MeV) $m_{\rm \tiny BARE}(\sigma)$ = 300 (dots), 400
(solid), 500 (dashes), 600 (long dashes) and 800 (dot-dashes).}
\label{SU2LsMsigmamass_fig}
\end{figure}

Using the partial wave amplitude Eqs. (\ref{pipi-partialwave}) and
(\ref{alphabeta}) it is straightforward to give a more detailed
discussion of the difficulty of regulating the infinity at $s= m_{\rm
BARE}^2 (\sigma)$ while still maintaining the good agreement near
threshold.  Consider replacing the denominator in
Eq. (\ref{pipi-partialwave}) according to the prescription
Eq. (\ref{conventionalreg}).  The effect of different constant widths $\Gamma$ in
Eq. (\ref{conventionalreg}) is illustrated in Fig. \ref{SU2LsMconstwidths_fig} for an arbitrary
choice of $m_{\rm BARE} (\sigma) = 560$ MeV.  It is seen that the
effect of increasing the width is to change the slope of the real part
of $T_0^0(s)$, $R_0^0(s)$
near threshold from positive to negative, which contradicts
experiment \cite{pipidata}.  Note that the unitarity bound $\left| R_0^0 (s) \right|
\le \frac{1}{2}$ is violated not too far away from threshold.
Theoretically, it is most natural to use instead of an arbitrary constant, the
``running'' perturbative width, 
\begin{equation}
\Gamma (s) = \frac{3}{16 \pi F_\pi^2 {\sqrt s}} {\sqrt {1 -
\frac{4m_\pi^2}{s}}} {\left[ m_{\rm BARE}^2 (\sigma) - m_\pi^2
\right]}^2,
\label{sigmawidth}
\end{equation}
as was tried also in \cite{AS}.  A plot of the real part of the
resulting amplitude $R_0^0(s)$ is shown in Fig. \ref{SU2LsMwidthsrun_fig} and is seen to badly
disagree with experiment.  This is due to the large
value of the perturbative width $\Gamma \left[ s=m_{\rm BARE}^2
(\sigma) \right]$.  It is amusing that the somewhat arbitrary
modification in the last factor of Eq. (\ref{sigmawidth}):
\begin{equation}
{\left[ m_{\rm BARE}^2 (\sigma) - m_\pi^2
\right]}^2 \longrightarrow {\left[ s - m_\pi^2
\right]}^2
\label{Amirwidth}
\end{equation}
greatly improves the agreement near threshold, as is also shown in
Fig. \ref{SU2LsMwidthsrun_fig}.  However somewhat beyond threshold the amplitude also starts
to deviate greatly from experiment.  Thus the prescription
Eq. (\ref{Amirwidth}) does not completely solve the problem, but may help fitting to experiment if the effects of
other possible particles are also taken into account.  The effect of
different values of $m_{\rm BARE} (\sigma)$ in Eq. (\ref{alphabeta})
is illustrated in Fig. \ref{SU2LsMsigmamass_fig} for this scheme.  

\subsection{K-Matrix unitarization}

We can force unitarity at all s for the scalar partial wave amplitude,
$T_0^0(s)$ by taking the tree amplitude ${T_0^0}_{\rm tree}(s)$ given
in Eqs. (\ref{pipi-partialwave}) and (\ref{alphabeta}) to coincide
with $K(s)$ in Eq. (\ref{regularization}).  To see what is happening
first consider ${T_0^0}_{\rm tree}(s)$ to be small (for example near
the $\pi \pi$ scattering threshold).  Then, in this single channel
case, 
\begin{equation}
S_0^0 \equiv 1 + 2iT_0^0 = 1 + 2i{T_0^0}_{\rm tree}(s) + ..., 
\end{equation}
so that $T_0^0(s)$ starts out as ${T_0^0}_{\rm tree}(s)$, which is a
reasonable approximation.  A presumably better approximation is
obtained by including more terms in an expansion of the denominator:
\begin{equation}
T_0^0(s)= \frac{{T_0^0}_{\rm tree}}{1 - i{T_0^0}_{\rm tree}} =
{T_0^0}_{\rm tree} \left[ 1 + i{T_0^0}_{\rm tree} +
{\left(i{T_0^0}_{\rm tree} \right)}^2 + ... \right].
\label{Tregexpan}
\end{equation}
As observed in \cite{AS} this has the structure of a bubble sum
in field theory.  However, in the present case one is working with the
partial wave, rather than the invariant, amplitude so there is no
integration over intermediate state momenta.  Of course, in either
case, crossing symmetry is lost.  While ${T_0^0}_{\rm tree}$ is gotten
from a crossing symmetric invariant amplitude, it is unlikely that the
specially iterated amplitude Eq. (\ref{Tregexpan}) can be gotten in
this way.  The advantage of the method is that it guarantees unitarity.
If, as is common, ${T_0^0}_{\rm tree}(s)$ starts getting too large,
$T_0^0$ will be chopped down to size.  For example if ${T_0^0}_{\rm
tree}$ gets very large:
\begin{equation}
S_0^0 \longrightarrow -1; \quad T_0^0 \longrightarrow i.
\label{reg_lim}
\end{equation}
The real part $R_0^0$ of $T_0^0$ vanishes in such a case while the
imaginary part $I_0^0 \longrightarrow 1$.  In particular this occurs,
as we see from Eq. (\ref{pipi-partialwave}), at the pole of
${T_0^0}_{\rm tree}$, where $s=m_{\rm BARE}^2$.  

With the tree level amplitude of Eq. (\ref{pipi-partialwave}), the
unitarized S-matrix takes the form:
\begin{equation}
S_0^0(s) = \frac { \left[ 1 + i\alpha (s) \right] \left[m_{\rm BARE}^2
(\sigma) - s \right] + i \beta (s) }{ \left[ 1 - i\alpha (s) \right]
\left[m_{\rm BARE}^2(\sigma) - s \right] - i \beta (s)},
\label{Salphabeta}
\end{equation}
where $\alpha (s)$ and $\beta (s)$ are given in
Eq. (\ref{alphabeta}).  Eq. (\ref{Salphabeta}) is sufficient for
comparing the predictions of the model [which contains the single
unknown parameter $m_{\rm BARE} (\sigma)$] with experiment.  However it is
also of interest to rewrite the amplitude so that it looks more like a
conventional resonance in the presence of a background.  Manipulating
Eq. (\ref{Salphabeta}) gives the factorized expression
\begin{equation}
S_0^0(s) = e^{2i\delta_{bg}(s)} \frac {{m^{\prime}}^2 (s) - s + i
\beta^\prime (s) }{{m^{\prime}}^2 (s) - s - i\beta^\prime (s) },
\label{Sfactorized}
\end{equation}
where,
\begin{eqnarray}
{\rm tan} \left[ \delta_{bg}(s) \right] &=& \alpha (s), \nonumber \\
{m^{\prime}}^2 (s) &=& m_{\rm BARE}^2(\sigma) +
\frac{\alpha(s)\beta(s)}{1 + \alpha^2(s)}, \nonumber \\
\beta^\prime (s) & = & \frac {\beta (s)}{1 + \alpha^2 (s)}.
\label{sreal}
\end{eqnarray}
This has the desired form although it should be noted that $m^\prime$
and $\beta^\prime$ are both $s$-dependent.  The T-amplitude which
follows from Eq. (\ref{Sfactorized}) and Eq. (\ref{regularization}) is
the sum of a background term and a modified resonance term
\begin{equation}
T_0^0(s) = e^{i\delta_{bg}(s)} {\rm sin} \delta_{bg} +
e^{2i\delta_{bg}(s)} \frac{\beta^\prime}{{m^{\prime}}^2 - s -
i\beta^\prime }.
\label{regT00}
\end{equation}

It is important to observe that the resonance mass and width (corresponding to a pole
in the complex s plane) are shifted from their bare values.  These new
values should be obtained from the complex solution{\footnote{For complex arguments the ln function in $\alpha (z)$ is chosen to have an imaginary piece lying between $-i\pi$ and $i\pi$.}}, $z_\sigma$ of:
\begin{equation}
{m^\prime}^2 \left( z \right)  - z - i \beta^\prime \left( z \right) =
0.
\label{poleequation}
\end{equation}  

We may choose to identify the {\it physical} mass and width of the
$\sigma$ from \footnote{A different definition of the resonance mass
and width was used in \cite{AS} but the numerical results are
close to each other.}

\begin{equation}
m_\sigma^2 - i m_\sigma \Gamma_\sigma = z_\sigma
\label{physicalmass}
\end{equation}

One should keep in mind that the resonance term is no longer
precisely of Breit-Wigner form.  

\begin{figure}
\centering
\epsfig{file=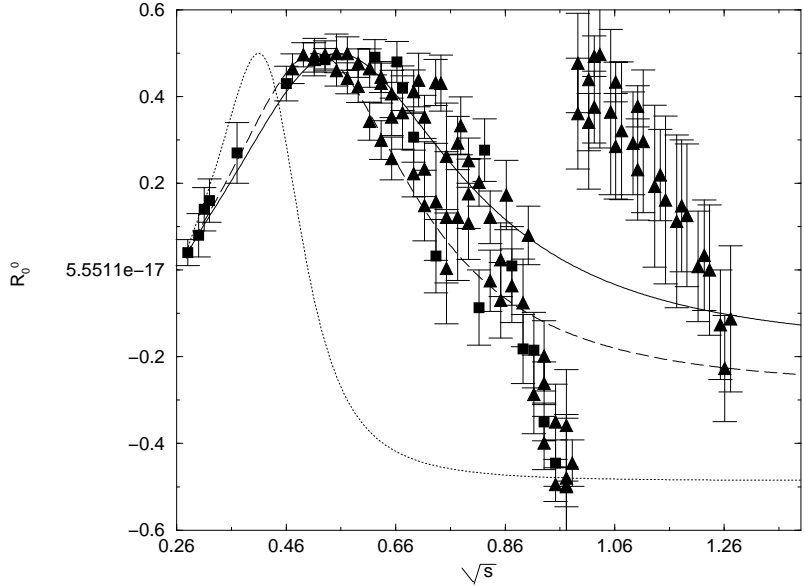, height=4in, angle=270}
\caption
{Comparison with experiment of Real part of the I=J=0 $\pi \pi$ scattering
amplitude in the SU(2) Linear Sigma Model, for $m_{\rm BARE} (\sigma)=0.5$ GeV (dots),
$m_{\rm BARE} (\sigma)=0.8$ GeV (dashes) and $m_{\rm BARE}(\sigma)=1$
GeV (solid).  Experimental data [44] are extracted from Alekseeva
{\it et al} (squares) and Grayer {\it et al} (triangles).}
\label{finalSU2LsMRamp}
\end{figure}

A plot of the real part $R_0^0 \left( s \right)$ of Eq. (\ref{regT00})
is presented in Fig. \ref{finalSU2LsMRamp} for the choices of (the single parameter in the
model) $m_{\rm BARE} \left( \sigma \right) = 0.5$ GeV, 0.8 GeV and 1
GeV.  It is seen, as already noted in \cite{AS}, that there is
reasonable agreement with experiment up to about ${\sqrt s} = 0.8$ GeV
if $m_{\rm BARE} \left( \sigma \right)$ lies in the 0.8 to 1 GeV
range.  Beyond ${\sqrt s} = 0.8$ GeV, the effects of the $f_0(980)$,
which does not appear in the two flavor model, are clearly important.
Also, the unitarized curves for $m_{\rm BARE} \left( \sigma \right)$
in the 0.8 to 1 GeV range give a reasonable looking description of the
threshold region, as opposed to the conventional unitarization scheme
of Eq. (\ref{conventionalreg}).  

\begin{figure}
\centering
\epsfig{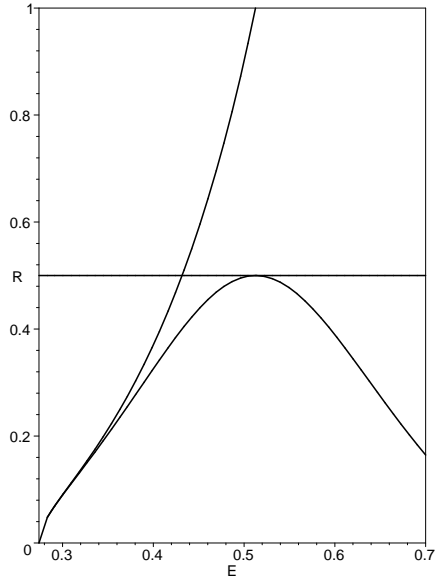}
\caption
{Comparison with experiment of Real part of K-matrix regularized I=J=0
$\pi
\pi$ scattering amplitude with the diverging real part of the tree approximation, for $m_{\rm BARE} (\sigma)=0.8$ GeV.}
\label{SU2tree_reg}
\end{figure}

Fig. \ref{SU2tree_reg} shows how the K-matrix unitarization works in detail by
comparing $R_0^0(s)$ with ${R_0^0(s)}_{tree}$.  It is seen that
${R_0^0}_{tree}$ already violates the unitarity bound at ${\sqrt
s}=0.43$ GeV.

Since we are regarding the K-matrix unitarization as a method of
approximating all the higher order corrections to the $\pi \pi$
scattering amplitude, it is clear that the quantities of physical
significance should not be the bare mass and width of $\sigma$ but
rather the pole mass $m_\sigma$ and width $\Gamma_\sigma$ defined
(with a usual convention) by Eqs. (\ref{poleequation}) and (\ref{physicalmass}).  These quantities were obtained numerically and
are given in Table 1 for the three choices of $m_{\rm BARE} \left(
\sigma \right)$ used above.  Evidently there are very substantial
shifts of the bare mass and the bare width.  The physical sigma pole
mass is around 0.45 GeV while the pole width is around 0.5-0.6 GeV for
$m_{\rm BARE} \left( \sigma \right)$ in the 0.8 to 1 GeV range.  

\begin{table}
\begin{tabular}{lccc}
$m_{\rm BARE}(\sigma)$ (GeV) & 0.5 & 0.8 & 1\\
$\Gamma_{\rm BARE} (\sigma )$ (GeV) &  0.311 & 1.58 & 3.22\\
$m_\sigma$ (GeV) & 0.421 & 0.458 & 0.449 \\
$\Gamma_\sigma$ (GeV) & 0.202 & 0.476 & 0.624 \\
$z_\sigma$ ${\rm GeV}^2$) &  0.177 - 0.085 i & 0.210 - 0.218 i & 0.202 - 0.281
i\\
$a_\sigma$ (${\rm GeV}^2$) & -0.015 + 0.078 i & 0.088 + 0.169 i & 0.158 + 0.188
i \\
$b_\sigma$ & -0.420 + 0.443 i & - 0.324 + 0.704 i & -0.274 + 0.753 i \\
\end{tabular}
\caption{Physical $\sigma$ parameters in the two flavor linear sigma model. }
\label{sigmatable}
\end{table}

In the present model we may qualitatively understand the decrease in
the $\sigma$ mass and also width by noting that $\alpha (s)$ and
$\beta (s)$ vary slowly with s.  If they are taken to be constant the
physical mass $m_\sigma$ would coincide with $m^\prime$ in
Eq. (\ref{sreal}) and the physical quantity $m_\sigma \Gamma_\sigma$
with $\beta^\prime$ in Eq. (\ref{sreal}).  Thus the negative sign of
the mass shift arises since the background piece of the amplitude
$\alpha (s)$ is negative.  A rather rough estimate may be made by
evaluating $\alpha (s)$ and $\beta (s)$ for $m_\pi = 0$ and $s$
small.  Then one finds $\beta \rightarrow \frac{3 m^4_{\rm
BARE}(\sigma)}{16 \pi {\rm F}_\pi^2}$ while $\alpha \rightarrow
-\frac{\beta}{m^2_{\rm BARE}(\sigma)}$.

It is interesting to note that our calculated amplitude $T_0^0 (s)$
can be reasonably well-approximated as 
\begin{equation}
T_0^0 (s) = \frac {a_\sigma}{s-z_\sigma} + b_\sigma,
\label{poleapprox}
\end{equation}
where the two complex numbers $a$ and $b$ are given in Table 1 for
different choices of $m_{\rm BARE} (\sigma )$.  Since this simple pole
dominated form reasonably fits experiment until the 700-800 MeV range it is not
surprising that various determinations of $m_\sigma$ and
$\Gamma_\sigma$ in the literature are roughly similar to the ones in
Table 1.  Often the $\sigma$ parameters are stated in terms of
$z^{\frac{1}{2}}$.  In the case where $m_\sigma = 0.458$ GeV we have
$z^{\frac{1}{2}} = 0.517 - i 0.240$ GeV.  This may be compared, for
example, with a treatment using a non-linear sigma model and including
the $\rho$ meson \cite{HSS1}.  That treatment gave a best fit for $z^{\frac{1}{2}}
= 0.585 - i 0.170$ GeV.  When it was refit \cite{HSS2} without the $\rho$ it
yielded $z^{\frac{1}{2}} = 0.493 - i 0.319$ GeV, which is closer to
the value in the present study (wherein, of course, spin 1 particles have not been 
included).

\section{Scattering in three flavor linear sigma models}

Here we study the pseudoscalar meson scattering amplitudes in the
three flavor linear sigma models discussed in the Introduction.  We
shall restrict attention to the J=0 elastic scattering amplitudes of 
\begin{eqnarray}
\pi \pi \longrightarrow \pi \pi, \nonumber \\
\pi K \longrightarrow \pi K, \nonumber \\
\pi \eta \longrightarrow \pi \eta,
\end{eqnarray}
which contain the scalars of the model - ($a_0$, $\kappa$, $\sigma$,
$\sigma^\prime$) - in the direct channel.  

Exactly the same K-matrix unitarization scheme of
Eqs. (\ref{regularization}) and (\ref{Kdef}) which was used in the two
flavor model will be employed.  In particular, no special assumptions
about the interplay of the $\sigma$ and $\sigma^\prime$ resonances in
$\pi \pi$ scattering will be made.  The tree amplitude will simply be
identified with $K$ in Eq. (\ref{regularization}).  The interesting
question about the treatment of $\pi \pi$ scattering is whether it can
fit the experimental data, given the complicated strong interferences
between the $\sigma$, $\sigma^\prime$ and contact term contributions.
The interesting question about the $\pi K$ scattering concerns the
properties of the $\kappa$ meson in the present model.  Finally the
$\pi \eta$  scattering is of methodological interest.  This is
because the well-established $a_0(980)$ resonance is expected to appear in a very
clean way, lacking interference from a strong contact term (or even
the possibility of potential interference when vector mesons are added
to the model), as explained, for example in \cite{BFS2}.  

We will first carry out the calculations using the standard
renormalizable form of the three flavor linear sigma model.  This is
characterized by the potential in Eq. (\ref{renorm_pot}).  Then the
whole model is extremely predictive!  After using as input the
well-established masses of the pseudoscalar nonet and pion decay
constant [Eq. (\ref{inputs})] there is only one quantity left to
choose in order to specify the scattering amplitudes.  This one
quantity may be taken to be the bare $\sigma$ mass, $m_{\rm \tiny
BARE} (\sigma)$.  The corresponding values of $m_{\rm \tiny
BARE} (\sigma^\prime)$ and $\theta_s$ are given in Fig. \ref{ms_mf}.
We shall also carry out the calculations for the most general chiral
symmetric potential.  This allows $m_{\rm \tiny BARE} (\sigma^\prime)$ 
and $\theta_s$ to be freely chosen, which is helpful for fitting
experiment.  As a possible justification for using a
non-renormalizable potential we mention that the model is an effective
one rather than the underlying QCD.  (It may be considered, for
example, to be a Wilson-type effective low energy Lagrangian.  
While non-renormalizable terms in the potential are technically
irrelevant they play a part in establishing the spontaneously broken
vacuum state and should be retained).  In any event the extra
parameters are being added in a chiral symmetric way.  

\subsection{$\pi \pi$ scattering}
The elastic amplitude for the three flavor linear sigma model in the
tree approximation was given in Eq. (\ref{pipiampl}) above.
Calculating the I=J=0 partial wave amplitude as in section IIA gives a
result which is a straightforward generalization of
Eq. (\ref{pipi-partialwave}):
\begin{equation}
T_{0{\rm tree}}^0 (s) = {\rm cos}^2 \psi \left[ \alpha (s) + \frac
{\beta (s)}{ m^2_{\rm \tiny
BARE} (\sigma) - s} \right] + {\rm sin}^2 \psi \left[ \tilde{\alpha} (s) + \frac
{\tilde{\beta} (s)}{ m^2_{\rm \tiny
BARE} (\sigma^\prime) - s} \right],
\label{pipi_SU3amp}
\end{equation}
where $\tilde{\alpha} (s)$ and $\tilde{\beta} (s)$ are respectively
gotten by replacing $m_{\rm \tiny BARE} (\sigma) \longrightarrow
m_{\rm \tiny BARE} (\sigma^\prime)$ in $\alpha (s)$ and $\beta (s) $ of
Eq. (\ref{alphabeta}).  The formula (\ref{pipi_SU3amp}) evidently
represents a sum of the $\sigma$ and $\sigma^\prime$ related contributions,
weighted by coefficients depending on the bare $\sigma-\sigma^\prime$
mixing angle $\psi$.  As before, we investigate the unitarized
amplitude based on Eq. (\ref{pipi_SU3amp}):
\begin{equation}
T_0^0(s) = \frac{T_{0{\rm tree}}^0(s)}{1 - iT_{0{\rm tree}}^0(s)},
\label{SU3reg}
\end{equation}
which is being interpreted as an approximation to including the
effects of the higher order corrections.  

\begin{figure}
\centering
\epsfig{file=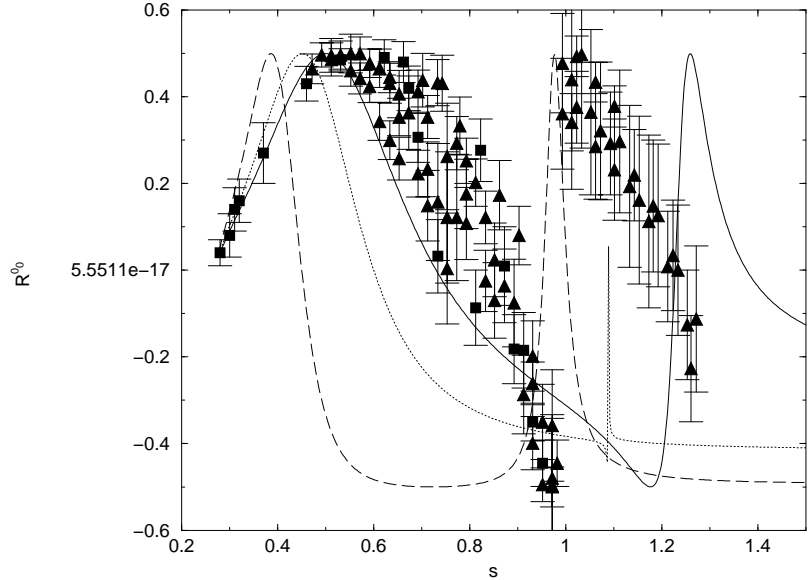, height=4in, angle=270}
\caption
{Comparison with experiment of Real part of the I=J=0 $\pi \pi$ scattering
amplitude in the renormalizable SU(3) Linear Sigma Model, for $m_{\rm BARE} (\sigma)=0.45$ GeV (dashes), $m_{\rm BARE} (\sigma)=0.6$ GeV (dots) and $m_{\rm BARE}(\sigma)=0.73$ GeV (solid).}
\label{SU3RLsMpipiamp}
\end{figure}

First, consider the renormalizable model.  To see how well it predicts
the interesting I=J=0 amplitude we may simply plot the real part of
Eq. (\ref{SU3reg}), $R_0^0(s)$ against $s$ for various choices of the
single undetermined parameter $m_{\rm \tiny BARE} (\sigma)$.  (Since
we are working in an elastic, exactly unitary approximation, the
imaginary part of the amplitude directly follows from the real part).
Clearly the effect of the $\sigma^\prime$ contribution is small when
the mixing angle $\psi$ is small.  Then one returns to the two flavor
case discussed in section II.  Recalling that $\psi \approx \theta_s +
54.7^o$ and referring to the second of Fig. \ref{ms_mf} shows that this will be the case when
$m_{\rm \tiny BARE} (\sigma)$ is around 0.6 GeV.  $R_0^0(s)$ for this
case is shown as the dotted line in Fig. \ref{SU3RLsMpipiamp}.  The experimentally derived points \cite{pipidata} are included for comparison.  This looks similar to the
curves for the two flavor case shown in Fig. \ref{finalSU2LsMRamp}, except that a sharp
blip appears at about 1.09 GeV corresponding to the $\sigma^\prime$
pole in $T_{0{\rm tree}}^0$.  Note, as can be seen from
Eq. (\ref{reg_lim}), that the real part $R_0^0(s)$ will vanish both at
$\sqrt {s} = m_{\rm \tiny BARE} (\sigma)$ and $\sqrt {s} = m_{\rm
\tiny BARE} (\sigma^\prime)$.  We saw in section IIB that the strong
interference with the background appreciably changes the position of
the physical pole in the complex $s$ plane so, while useful for
understanding Fig. \ref{SU3RLsMpipiamp}, these zeroes of $R_0^0$ do not give true
parameterizations of the pole position.  

We have already learned that the choice $m_{\rm \tiny BARE} ( \sigma )
= 0.6$ GeV is too low for a good fit to $R_0^0(s)$ in the region of
$\sqrt s$ up to about 0.6 GeV.  Increasing $m_{\rm \tiny BARE} (
\sigma )$ improves the fit to this region and also allows the effects
of the $\sigma^\prime$ to come into play.  This is shown as the solid
line in Fig. \ref{SU3RLsMpipiamp} which corresponds to the choice $m_{\rm \tiny BARE} (
\sigma ) = 0.73$ GeV.  Unfortunately (as expected from Fig. \ref{finalSU2LsMRamp}) this
choice is still not high enough for a good fit in the low energy
region.  Furthermore, the structure which should correspond to the
experimental $f_0(980)$ resonance has been pushed too high.
Increasing $m_{\rm \tiny BARE} ( \sigma )$ further will, as the first
of Fig. \ref{ms_mf} shows, push this structure even higher.  One still must check
to see if lowering $m_{\rm \tiny BARE} ( \sigma )$ below 0.6 GeV can
work.  The dashed line in Fig. \ref{SU3RLsMpipiamp} shows $R_0^0(s)$
for  $m_{\rm \tiny BARE} (\sigma ) = 0.45$ GeV.  In this case the
structure near the $f_0(980)$ is in the right place but much too
narrow.  The structure at lower energies is also very badly
distorted.  

By examining the evolution with increasing $m_{\rm \tiny BARE} (\sigma
)$ of the three curves in Fig. \ref{SU3RLsMpipiamp}, one sees that it
is not possible to get a good fit for any choice of $m_{\rm \tiny BARE} (\sigma
)$.  Nevertheless the qualitative prediction of $R_0^0(s)$ is clearly
a sensible one.  It is therefore tempting to see if there is an easy
way to fix up the fit.  

In a sense, the difficulty in obtaining a good fit arises because only
one parameter - taken to be $m_{\rm \tiny BARE} ( \sigma )$ - is
available for adjustment to give agreement with a rather complicated
experimental shape.  The easiest way to proceed is to modify some
parameters involved in the calculation.  If a parameter to be varied
is arbitrarily chosen there is however a danger of breaking the chiral
symmetry relations intrinsic to the model.  For example, suppose we
choose to vary the coupling constant of the bare sigma to two pions.
This three-point coupling constant, as mentioned in section I, is
related to the masses of the particles involved (see Appendix).  Then,
changing it without changing the masses will break the underlying
chiral symmetry.  Of course, we have written the formula for the tree
ampliutde, Eq. (\ref{pipiampl}) in such a way that the correct
relations for the 4 and 3 point coupling constants are automatically
taken into account for any choice of the contained parameters $m_{\rm
\tiny BARE} ( \sigma )$, $m_{\rm \tiny BARE} ( \sigma^\prime )$ and
$\theta_s$.  In fact these three parameters may be freely chosen in
the linear sigma model Eq. (\ref{LsMLag}) with an arbitrary chiral
invariant potential, $V_0$.  It is only by restricting $V_0$ to be
renormalizable that one can relate $m_{\rm \tiny BARE} (
\sigma^\prime)$ and $\theta_s$ to  $m_{\rm \tiny BARE} ( \sigma )$.
Thus we can freely vary  $m_{\rm \tiny BARE} (\sigma^\prime)$ and 
$\theta_s$ in addition to $m_{\rm \tiny BARE} ( \sigma )$ if we choose
to obtain the tree amplitude from the non-renormalizable model of
Eq. (\ref{LsMLag}).  

\begin{figure}
\centering
\epsfig{file=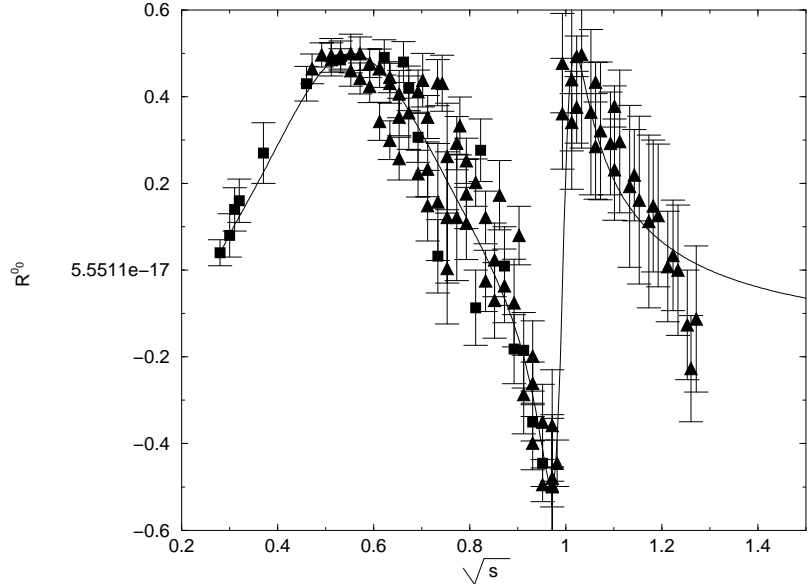, height=4in, angle=270}
\caption
{Comparison of our best fit for the Real part of the I=J=0 $\pi \pi$ scattering
amplitude in the non-renormalizable SU(3) Linear Sigma Model with experiment.}
\label{finalSU3LsMNRpipiamp}
\end{figure}

In such a model it is easy to fit the experimental data for
$R_0^0(s)$.  A best fit obtained using the MINUIT package is shown in
Fig. \ref{finalSU3LsMNRpipiamp}.  It corresponds to the parameter
choices $m_{\rm \tiny BARE} (\sigma) = 0.847$ GeV, $m_{\rm \tiny BARE}
(\sigma^\prime) = 1.30$ GeV and $\psi = 48.6^o$.  The physical masses
and widths are obtained, as in Eq. (\ref{physicalmass}) in the two
flavor case, from the pole positions in the complex $s$ plane.  These,
together with the residues at the poles, are listed in Table
\ref{SU3pipiTable}.
\begin{table}
\begin{tabular}{lcc}
  & $\sigma$ & $\sigma^\prime$ \\ \tableline
$m_{\rm \tiny BARE}$ (GeV) &  0.847 & 1.300 \\
$\Gamma_{\rm \tiny BARE}$ (GeV) & 0.830 & 4.109  \\
$m$ (GeV) & 0.457 & 0.993 \\
$\Gamma$ (GeV) & 0.632 & 0.051 \\
$z$ (${\rm GeV}^2$) &  0.209 - i 0.289 & 0.986 - i 0.051 \\
$a$ (${\rm GeV}^2$) & 0.167 + i 0.210 & 0.053 - i 0.005  \\ \tableline
$b$ & -0.248 +0.856 i\\
\end{tabular}
\caption{Physical $\sigma$ and $\sigma^\prime$ parameters obtained
from best fit using non-renormalizable SU(3) linear sigma model.}
\label{SU3pipiTable}
\end{table}

For orientation we first note that the parameters describing the lower
mass scalar, $\sigma$ are in the same range, as expected, as the
parameters of Table \ref{sigmatable} which give good fits to the low energy data
using the $\sigma$ in the two flavor case.  In fact the masses are
very close to each other but the effect of the additional flavor
requires a somewhat greater width parameter.  The contribution of the
$\sigma$-pole to $T_0^0(s)$ is read off as:
\begin{equation}
\frac{a_\sigma}{s-z_\sigma} = \frac{-a_\sigma}{m_{\sigma}^2 - s - i
m_\sigma \Gamma_\sigma} = \frac {-0.167 - i 0.210}{0.209 - s - i
0.289}.
\end{equation}

Note that this form is very different from a pure Breit-Wigner form
which would require the numerator to be $0.289$ ${\rm GeV}^2$.  This
illustrates (as does the two flavor case) the importance of the
interplay between the resonance and the ``background''.  It also
illustrates the possible difficulty of trying to get properies of the
$\sigma$ from experiment with the use of a pure Breit Wigner
approximation.  Note (again) from Table \ref{SU3pipiTable} that the
physical mass and width, $m_\sigma$ and $\Gamma_\sigma$ respectively,
are very much reduced from their bare values.  The detailed mechanism
is evidently similar to what we described in section II for the two
flavor case.  It seems reasonable to consider the physical values of
$m_\sigma$ and $\Gamma_\sigma$ to be the ones which are significant. 

The well-established resonance $f_0(980)$ will be identified with the
$\sigma^\prime$.  The contribution of the $\sigma^\prime$ pole to
$T_0^0(s)$ is read off as:
\begin{equation}
\frac{a_{\sigma^\prime}}{s-z_{\sigma^\prime}} = \frac{-a_{\sigma^\prime}}{m_{\sigma^\prime}^2 - s - i
m_{\sigma^\prime} \Gamma_{\sigma^\prime}} = \frac {-0.053 - i 0.005}{0.986 - s - i
0.051}.
\end{equation}

In this case the form of a pure Breit-Wigner would require that the
numerator be $+0.051$.  To a reasonable approximation this holds
except for an overall sign.  Now reference to the formula
Eq. (\ref{regT00}) for a Breit Wigner with a background, shows that
the background phase $\delta_{\rm bg} = \frac{\pi}{2}$ must be supplying this negative sign.  Clearly the negative sign is required by the
experimental data showing the real part $R_0^0(s)$ to be negative
before and positive after the resonance.  It was noted \cite{HSS1}
that this is an example of the well-known Ramsauer-Townsend effect in
scattering theory.  It is also interesting to observe from Table
\ref{SU3pipiTable} that the bare mass of the $\sigma^\prime$ is
substantially shifted down from $1.300$ GeV (where a zero of
$R_0^0(s)$ remains, as previously discussed) to about 1 GeV.  The bare
width is even more substantially shifted from about 4 GeV to 50 MeV!

\begin{figure}
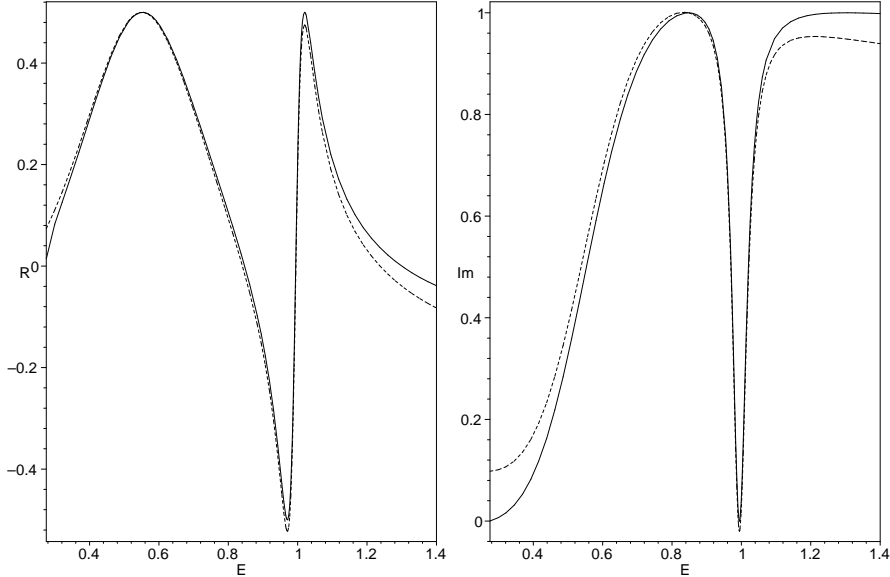

\centering
\epsfig{file=./finalSU3NRLsMpipiRepole.eps, height=3in, angle=0}
\epsfig{file=./finalSU3NRLsMpipiImpoleshifted.eps, height=3in, angle=0}
\caption
{Comparison of (left) Real and (right) Imaginary parts of pole approximation
Eq. (\ref{SU3pipipoleapprox}) (dashed line) with those of our predicted
amplitude Eq. (3.2) (solid line).}
\label{pipipoleapprox}
\end{figure}

One might wonder whether the simple pole dominance approximation
Eq. (\ref{poleapprox}) for the two flavor case can be generalized to
this more complicated three flavor case containing two poles.  It
turns out to be true;  the prediction of our model can be numerically
approximated by the sum of the two pole terms and a suitably chosen
constant:
\begin{equation}
T_0^0(s) = R_0^0(s) + i I_0^0(s) \approx \frac{a_\sigma}{s-z_\sigma} +
\frac{a_{\sigma^\prime}}{s-z_{\sigma^\prime}} + b,
\label{SU3pipipoleapprox}
\end{equation}
where the numbers $a_\sigma$, $a_{\sigma^\prime}$, $z_\sigma$,
$z_{\sigma^\prime}$ and b are listed in Table \ref{SU3pipiTable}.
This is illustrated in Fig. \ref{pipipoleapprox} for $R_0^0(s)$ and $I_0^0(s)$.  In these figures Eq. (\ref{SU3pipipoleapprox}) is being
compared with our prediction in Fig. \ref{finalSU3LsMNRpipiamp}.  Of
course, there is no reason to use Eq. (\ref{SU3pipipoleapprox})
instead of the more accurate and complicated formula
Eq. (\ref{SU3reg}) but it nicely shows the dominating effect of
the poles.  The pole approximation is seen however not to be very
accurate near threshold.  

Incidentally the deep dip in $I_0^0(s)$ at what we have found to be
the $\sigma^\prime$ physical pole position also represents the
Ramsauer-Townsend effect.  This appears in $R_0^0(s)$  as a
``flipped'' resonance curve, as discussed above.  Actually this
Ramsauer-Townsend phenomenon can be pictured in an alternative manner.
If we consider $R_{0{\rm tree}}^0(s)$ corresponding to two ``bare''
resonances, one following the other, we see that there must be a point
in between them where $R_{0{\rm tree}}^0 = 0$.  Then
Eq. (\ref{SU3reg}) shows that, after K-matrix unitarization, $R_0^0$
will also vanish at this point.  This point appears visually as the
pole position zero of a ``flipped'' standard
resonance curve!  In the Ramsauer-Townsend interpretation the flipping
is interpreted as a background phase of $\frac{\pi}{2}$.  Our explicit
determination of the pole positions for the sigma model amplitude
shows that this is the pole which captures the dynamics of the
$f_0(980)$.  Its narrow width is seen to be the result of its getting
``squeezed'' between two nearby ``bare'' poles by the unitarization in
this model.  

\subsection{$\pi$ K scattering}

We are interested in the I=1/2, J=0 scattering amplitude in order to
investigate the properties of the $\kappa$ resonance in the direct
channel.  The tree level amplitude involves $\kappa$ exchanges in the
$s$ and $u$ channels, $\sigma$ and $\sigma^\prime$ exchanges in the
$t$ channel as well as a four point contact term.  The relevant tree
level invariant amplitude may be written as:
\begin{equation}
A^{1/2} \left( s,t,u \right) = - g_{K}^{(4)} +\frac{3}{2} 
\frac{g_{\kappa K\pi}^2}{m^2_{\rm \tiny BARE}(\kappa) - s} - \frac{1}{2} \frac{g_{\kappa K
\pi}^2}{m^2_{\rm \tiny BARE}(\kappa) - u} - \frac{g_{\sigma\pi\pi} g_{\sigma K {
K}}}{m^2_{\rm \tiny BARE}(\sigma) - t} - \frac{g_{\sigma^\prime \pi\pi} g_{\sigma^\prime K {
K}}}{m^2_{\rm \tiny BARE}(\sigma^\prime) - t},
\label{piKtree}
\end{equation}
where $s$, $t$ and $u$ are the usual Mandelstam variables.  The four
point contact interaction $g_K^{(4)}$ and the bare three point
coupling constants shown are listed in the Appendix.  As for the cases
of $\sigma$ and $\sigma^\prime$ we have put a subscript BARE on the
$\kappa$ mass to indicate that the location of the physical pole after
unitarization may come out different from this.  The scalar partial
wave tree amplitude is next defined by
\begin{equation}
T_{0{\rm tree}}^{1/2} =  \rho(s) \int^1_{-1} d{\rm cos} \theta \, A^{1/2} (s,t,u).
\label{piKpartialwave}
\end{equation} 
Note that $\rho (s)$ was already defined by
Eqs. (\ref{kinematicalfactor}) and (\ref{cofm}). The specific formula
for Eq. (\ref{piKpartialwave}) in the present model is a bit lengthy and is shown in the
Appendix.

According to our plan we do not introduce any new parameters for
unitarization and simply write
\begin{equation}
T_0^{1/2} = \frac{T_{0{\rm tree}}^{1/2}}{ 1 - i T_{0{\rm
tree}}^{1/2}},
\label{piKunitarized}
\end{equation}
which is related to the corresponding S-matrix element by
Eq. (\ref{regularization}).  

As mentioned in the Introduction the value of $m_{\rm \tiny
BARE}(\kappa)$ is independent of whether or not the chiral invariant
potential in Eq. (\ref{LsMLag}) is renormalizable, but depends only
on the set of input parameters [e.g. Eq. (\ref{inputs})].  This may be
seen from the equation
\begin{equation}
m^2_{\rm \tiny BARE}(\kappa) = \frac{F_K m_K^2 - F_\pi m_\pi^2}{F_K -
F_\pi}
\label{barekappamass}
\end{equation}
which follows from Eqs. (\ref{decayconstants}), (\ref{piKmasses}) and
(\ref{scalarmasses}) in the isotopic spin invariant limit.  This means
that there are no new unknown quantities beyond those used in the fit
to the $\pi \pi$ scattering amplitude above.  However we observe that
the predicted value of $m_{\rm \tiny BARE}(\kappa)$ is very sensitive
to the difference $F_K - F_\pi$.  Actually the choice of input
parameters given in Eq. (\ref{inputs}) results in a somewhat too high
prediction for $F_K$, as mentioned before.  

One might therefore wonder
whether the choice of input parameters in Eq. (\ref{inputs}) unfairly
biases our treatment of $\pi K$ scattering by giving a too small value
for $m_{\rm \tiny BARE}(\kappa)$.  In order to check this we will also consider the slightly different choice
of input parameters {\footnote{We then predict $m_\eta \approx 0.53$ GeV rather than the experimental value of 0.547 GeV}}$(m_\pi, m_K, m_{\eta^\prime},
F_\pi, F_K)$.  This will not affect the $\pi \pi$ scattering results
in the non-renormalizable model just discussed.  We first choose $F_K = 1.16
F_\pi$ which is slightly smaller than the physical value but has the
advantage that it gives $m_{\rm \tiny BARE}(\kappa) = 1.3$ GeV which
yields a zero for $T_0^{1/2}(s)$ at 1.3 GeV, in agreement with the
experimental data.  For this new input set we also have explicitly
checked that there is still no possibility of getting a good fit to
$\pi \pi$ scattering in the renormalizable model.  

\begin{figure}
\centering
\epsfig{file=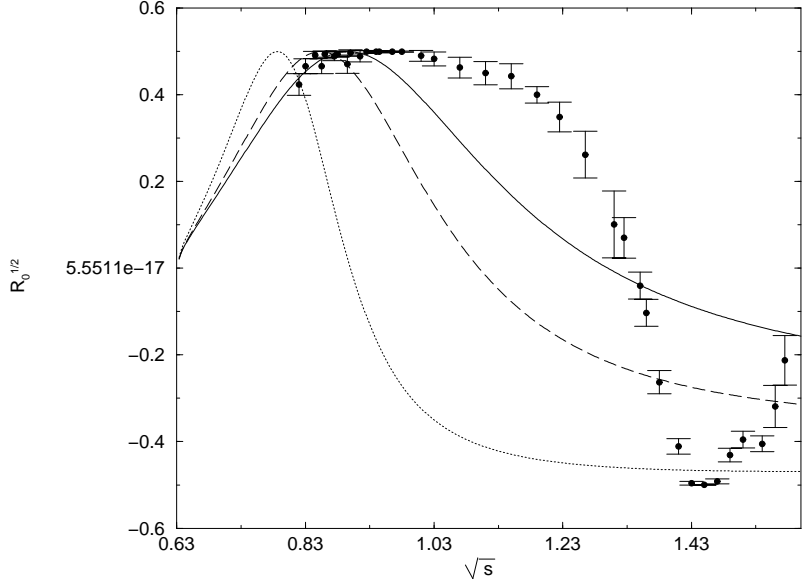, height=4in, angle=270}
\caption
{Comparison of our prediction for the Real part of the
I=$\frac{1}{2}$, J=0 $\pi K$ scattering
amplitude in the non-renormalizable SU(3) Linear Sigma Model with
experiment.  The curves correspond to $m_{\rm \tiny BARE} (\kappa) =
1.3$ GeV (solid), 1.1 GeV (dashed) and 0.9 GeV (dotted).  The
experimental data are extracted from [45].}
\label{finalSU3LsMpikNR}
\end{figure}

With this choice of input and other coupling constants taken in
agreement with those found in the best fit to $\pi \pi$ scattering we
have the prediction for the real part of the amplitude $R_0^{1/2}(s)$
shown in Fig. \ref{finalSU3LsMpikNR}.  The experimental data \cite{Aston}, which start around 0.83 GeV
and go to about 1.6 GeV are also shown in this graph.  It is seen that
the prediction from the linear sigma model agrees with the data from
about 0.83 GeV to about 0.92 GeV.  However at higher energies the
predicted curve lies much too low until about 1.35 GeV and thereafter
seems to completely miss the structure which is usually associated
with the $K_0^*(1430)$ resonance.  

Fig. \ref{finalSU3LsMpikNR} also shows the predictions for the cases when $m_{\rm \tiny
BARE}(\kappa) = 1.1$ GeV (corresponding to $F_K$ taking its
experimental value) and $m_{\rm \tiny BARE}(\kappa) = 0.9$ GeV
(corresponding to the input choice of Eq. (\ref{inputs}).  These are
in worse agreement with experiment and also seem to miss the
$K_0^*(1430)$ structure.  

As in the two flavor $\pi \pi$ case, which also contains only a single
direct channel resonance we have found that the predicted amplitude is
fairly well approximated as the sum of a pole term and a constant:
\begin{equation}
T_0^{1/2}(s) \approx \frac{a_\kappa}{s-z_\kappa} + b_\kappa.
\label{piKpoleapprox}
\end{equation}

The values for $z_\kappa$, $a_\kappa$ and $b_\kappa$ corresponding to
the three different choices of input parameters are shown in Table
\ref{piKTable}.  Again we identify the physical mass and width by 
\begin{equation}
m_\kappa^2 - i m_\kappa \Gamma_\kappa = z_\kappa.
\label{kappaphysical}
\end{equation}
It is notable that the pole position mass is always close to 800 MeV
regardless of the choice of $m_{\rm \tiny BARE}(\kappa)$.  Furthermore
the widths obtained from Eq. (\ref{kappaphysical}) are substantially
reduced from their ``bare'' (tree level) values, but are more
sensitive to the choice of $m_{\rm \tiny BARE}(\kappa)$.

All in all, the properties of the $\kappa$ obtained here are very
analogous to those of the $\sigma$ in either the two or three flavor
treatments of $\pi \pi$ scattering.  Compare with Fig. \ref{finalSU2LsMRamp}, for example.

It does seem that the pole mass, Eq. (\ref{kappaphysical}), of the
$\kappa$ is a good indication of the energy region where it provides a
reasonable fit to the data.  It also seems clear that the physics
associated with the higher mass $K_0^*(1430)$ is not being taken into
account in this model.  

\begin{table}
\begin{tabular}{lccc}
$m_{\rm \tiny BARE}(\kappa)$ (GeV) &  0.9 & 1.1 & 1.3 \\
$\frac{F_K}{F_\pi}$, $\theta_p$ & 1.4,$2.5^o$ & 1.23,$-4.6^o$ & 1.16,$-8.8^o$ \\
$\Gamma_{\rm \tiny BARE}(\kappa)$ (GeV)& 0.403 & 1.138 & 2.35  \\
$m_\kappa$ (GeV) & 0.799 & 0.818 & 0.798 \\
$\Gamma_\kappa$ (GeV) & 0.257 & 0.461 & 0.614 \\
$z_\kappa$ (${\rm GeV}^2$) &  0.639 - i 0.205 & 0.669 - i 0.378 &
0.637 - i 0.490 \\
$a_\kappa$ (${\rm GeV}^2$) & -0.043 + i 0.190 & 0.096 + i 0.340 &
0.263 + i 0.378  \\ 
$b_\kappa$ & -0.438 + i 0.420 & -0.419 + i 0.660 & -0.357 + i 0.800\\
\end{tabular}
\caption{Physical $\kappa$ parameters obtained in the non-renormalizable
SU(3) linear sigma model for different values of $m_{\rm \tiny BARE}(\kappa)$ which result from different choices of input parameters.}
\label{piKTable}
\end{table}

\subsection{$\pi \eta$ scattering}

\begin{figure}
\centering
\epsfig{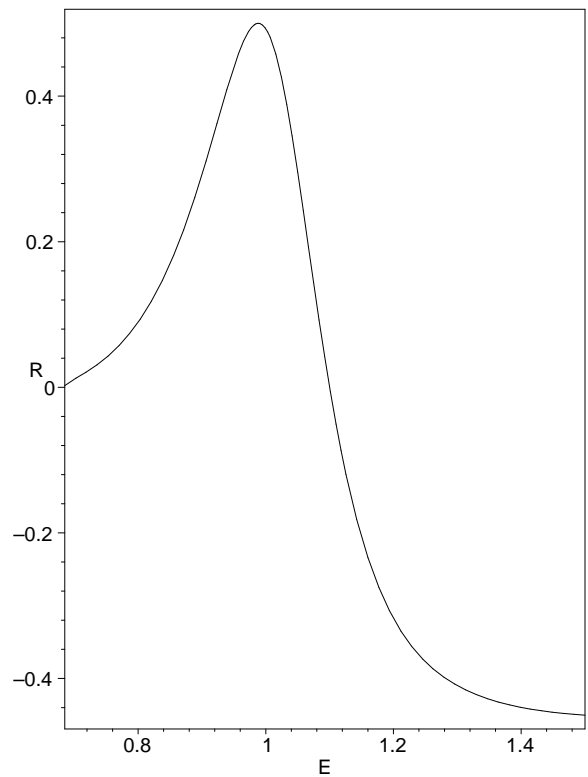}
\caption
{Prediction for the Real part of the I=1, J=0 $\pi \eta$ partial wave scattering
amplitude Eq. (\ref{pieta_unitarization}) in the non-renormalizable SU(3) Linear Sigma Model
(parameters as in second
column of Table \ref{pieta_table}).}
\label{finalSU3LsMpieta}
\end{figure}

The tree level invariant amplitude takes the form 
\begin{eqnarray}
A^{1}(s,t,u) &=&  - g_{\eta}^{(4)} + g_{a_0 \pi \eta}^2 \left[\frac{1}
{m^2_{\rm \tiny BARE}(a_0) - s} 
+\frac{1}{m_{\rm \tiny BARE}^2(a_0) - u}\right] \nonumber \\
&+& \frac{g_{\sigma\pi\pi}
g_{\sigma\eta\eta}}{m_{\rm \tiny BARE}^2(\sigma) - t} + \frac{g_{\sigma^\prime\pi\pi} g_{\sigma^\prime
\eta\eta}}{m_{\rm \tiny BARE}^2(\sigma^\prime) - t},
\label{pieta_treeamp}
\end{eqnarray}
where the four point contact term $- g_{\eta}^{(4)}$ as well as the
three point coupling constants are listed in the Appendix.  Other
conventions are the same as above.  Similarly the scalar partial wave
amplitude is
\begin{equation}
T_{0 tree}^1  =  \rho(s) \int^1_{-1} d{\rm cos} \theta \, A^1 (s,t,u),
\label{pieta_partialwave}
\end{equation} 
which is also listed in the Appendix.  Again we unitarize by
substituting this into the formula
\begin{equation}
T_0^1 = \frac{T_{0 tree}^1}{1 - i T_{0 tree}^1}.
\label{pieta_unitarization}
\end{equation}
Since there is apparently no experimental phase shift analysis
available for this channel, we will have to be content to just present
our theoretical results and compare with the mass and width of the
experimental $a_0(980)$ resonance.  It was already noted that the
renormalizable model [with the input Eq. (\ref{inputs})] yields the
somewhat too low bare mass (which gets shifted down by unitarization) of 913 MeV.  We
will also present the results for the non-renormalizable model which
gave a good picture of $\pi \pi$ scattering and for which we are still
free to choose $m_{\rm \tiny BARE}(a_0)$.  A value  $m_{\rm \tiny
BARE}(a_0) = 1.100$ GeV gives roughly the correct ``physical mass''
and the plot of the real part of Eq. (\ref{pieta_unitarization}) for
this choice is shown in Fig. \ref{finalSU3LsMpieta}.  The result of the
regularization is generally similar to the curves obtained for the
$\sigma$ in $\pi \pi$ scattering and the $\kappa$ in $\pi K$
scattering.  We have found in this case too that the predicted
amplitude is reasonably well approximated by the sum of a pole and a
constant:
\begin{equation}
T_0^1(s) = \frac{a_{a_0}}{s - z_{a_0}} + b_{a_0}.
\end{equation}
The physical mass and width are found from 
\begin{equation}
m_{a_0}^2 - i m_{a_0}\Gamma_{a_0} = z_{a_0}
\end{equation}
and the appropriate values for the two cases mentioned are listed in
Table \ref{pieta_table}.  While Fig. \ref{finalSU3LsMpieta} seems to be
just what one would expect for the real part of a resonance amplitude,
Table \ref{pieta_table}, as in the previous cases, reveals some
interesting features.  First, since $a_{a_0}$ is clearly different
from $-{\rm Im}(z_{a_0})$, the resonance is not a pure Breit Wigner.
The location of the physical pole is close to the positive peak of
$R_0^1(s)$ rather than to its zero, as would hold for a Breit Wigner.
Compared to the scalar resonance $\pi \pi$ and $\pi K$ channels we
notice that there are smaller shifts going from $m_{\rm \tiny
BARE}(a_0)$ to $m_{a_0}$ and from $\Gamma_{\rm \tiny
BARE}(a_0)$ to $\Gamma_{a_0}$.  This is reasonably interpreted as due
to less effect of interference with the background.  This is manifest
in the non-linear sigma model approach to $\pi \eta$ scattering
\cite{BFS2} and can thus be understood as a consequence of the
similarity of the non-linear and linear chiral models.  In addition,
we note that $\Gamma_{a_0}$ is predicted to be somewhat larger than
the experimental value \cite{PDG} of 50-100 MeV.  Nevertheless, the prediction is
qualitatively reasonable.

\begin{table}
\begin{tabular}{lcc}
$m_{\rm \tiny BARE}(a_0)$ (GeV) &  0.913 & 1.100 \\
$\Gamma_{\rm \tiny BARE}(a_0)$ (GeV)& 0.129 & 0.381 \\
$m_{a_0}$ (GeV) & 0.890 & 1.013 \\
$\Gamma_{a_0}$ (GeV) & 0.109 & 0.241 \\
$z_{a_0}$ (${\rm GeV}^2$) &  0.793 - i 0.097 & 1.027 - i 0.244 \\
$a_{a_0}$ (${\rm GeV}^2$) & -0.065 + i 0.064 & - 0.076 + i 0.200   \\ 
$b_{a_0}$ & -0.299 + i 0.204  & -0.312 + i 0.408\\
\end{tabular}
\caption{Physical $a_0$ parameters in renormalizable (first column)
and non-renormalizable (second column) SU(3) linear sigma model using
corresponding best-fit parameters from $\pi \pi$ scattering.}
\label{pieta_table}
\end{table}

\section{Summary and Discussion}

We have treated the three flavor linear sigma model as a ``toy model''
for examining the role of possible light scalar mesons in the $\pi
\pi$, $\pi K$ and $\pi \eta$ scattering channels.  This is a highly
predictive model which contains only one free parameter, which may to
taken as $m_{\rm \tiny BARE}(\sigma)$, in the renormalizable case.  If
we give up renormalizability for this effective Lagrangian but
maintain chiral symmetry in a straightforward way, $m_{\rm \tiny
BARE}(\sigma^\prime)$, the scalar mixing angle $\theta_s$ and $m_{\rm \tiny BARE}(a_0)$ may
also be freely chosen, which is helpful for fitting experiment in the
desired energy range of threshold to the 1+ GeV region.  Our approach
just involves computing the tree amplitude for each channel and
unitarizing by a simple ``K-matrix'' prescription which does not
itself introduce any new parameters.  In general the unitarization has
very important effects converting ``bare'' scalar meson masses and
widths into ``physical'' ones.  It turns out that there is not too
much ``wiggle room'' in this procedure so that what results is
characteristic of the model (and the unitarization scheme).  This
tightness comes from the demand that the starting tree amplitudes
satisfy chiral symmetry restrictions.  This means, as discussed in
section I, that the four point contact interaction vertices are
related to the three point interactions which are related to the
particle masses (two point objects) which are related in turn to the
one point terms (pseudoscalar meson decay constants).  We chose the
inputs to be the four pseudoscalar masses and the pion decay constant
[Eq. (\ref{inputs})].  However the pseudoscalar mixing angle and kaon
decay constant were not perfectly predicted so there is already a
source of error present before even going to the scattering
amplitudes.  Nevertheless we investigated this point by considering an
alternative input set obtained by using $F_K$ instead of $m_\eta$ and
found that there was not much qualitative change for the scattering
predictions.  

Our point of view in this paper is to see what are the results of
computing in a relatively simple and natural model for the purpose of
comparison with other (and possibly future) more elaborate
treatments.  It seems to us that the results are interesting and
instructive.  In the simpler two flavor case, which was applied in
\cite{AS} to a lower energy treatment of $\pi \pi$ scattering,
the results were already reasonable.  Here we have, in section II,
reviewed the two flavor case in a slightly different way as
preparation for the more complicated three flavor case.  We have also
made some new comments and suggested an alternative ``naive''
unitarization procedure which might be handy for future studies.  

Table \ref{summary_table}  contains a brief summary of the physical
masses and widths of the scalar mesons predicted in the present model
and discussed in some detail in section III.  In the cases of the
$f_0(980)$ and $a_0(980)$ resonances comparison is being made with
experimental values \cite{PDG}.  In the cases of the $\sigma$ and
the $\kappa$, which are less well-established experimentally, we have
compared with the earlier computations of the Syracuse group
\cite{HSS1,BFSS1,BFSS2} which were based on a non-linear chiral
effective Lagrangian treatment, including vector mesons.  Many other
authors \cite{Other} were led to similar predictions for the $\sigma$ while
a similar prediction for the $\kappa$ was made by \cite{Ishida}.

\begin{table}
\begin{tabular}{lcccc}
 &  $\sigma$ & $f_0$ & $\kappa$  & $a_0$ \\ \tableline
Present Model& & & & \\ 
mass (MeV), width (MeV) & 457, 632 & 993, 51 & 800, 260-610 & 890-1010, 110-240 \\
Comparison & & & & \\
mass (MeV), width (MeV)& 560, 370 & 980$\pm 10$, 40-100 &
900, 275 & 985,  50-100 \\ 
\end{tabular}
\caption{Predicted ``physical'' masses and widths in MeV of the nonet
of scalar mesons contrasted with suitable (as discussed in the text)
comparison values.}
\label{summary_table}
\end{table}

The predicted properties of the $\sigma$ and $f_0$ in the present
model come from their role in $\pi \pi$ scattering as discussed in
section III A.  It was found that the single parameter describing the
renormalizable model could not be adjusted to give a reasonable fit to
the experimental data.  This could be done when the renormalizability
condition was relaxed.  Neither the physical $\sigma$ nor the physical
$f_0$ are described by simple Breit Wigner terms.  Both have masses
and widths greatly reduced from their ``bare'' values by the
unitarization procedure.  The light, broad $\sigma$ is somewhat
lighter and broader than the comparison one obtained in the non-linear
model \cite{HSS1}.  (However when the vector meson contribution in the non-linear
model was consistently eliminated \cite{HSS2} the $\sigma$ in that
model also became broader and lighter).  The $f_0$ obtained
approximately looked like a Breit-Wigner in a background which has a
phase $\delta_{bg} = \frac{\pi}{2}$.  This is known as the
Ramsauer-Townsend effect in scattering theory.  The fact that it
emerges in the present model was noted to be explicable in terms of
the region between two neighboring ``bare'' resonances getting
squeezed by unitarization.

The entries in Table \ref{summary_table} for the $\kappa$ mass and
width require some explanation.  The bare $\kappa$ mass and width in
this model are uniquely predicted once the input parameters are
specified, regardless of whether or not the potential is taken to be
renormalizable.  However the predictions of the $\kappa$ parameters
are very sensitive to $F_K$ (which measures the deviation of the
vacuum from exact SU(3) flavor symmetry in this model).  Thus we
allowed different input sets yielding different bare $\kappa$ masses,
as discussed in section IIIB.  Whatever reasonable choice was made,
the unitarization always brought the physical $\kappa$ mass down to
around 800 MeV.  However the physical width is more dependent on this
choice.  Furthermore, as shown in Fig. \ref{finalSU3LsMpikNR}, the $\kappa$ resonance can
only explain the lower energy $\pi K$ scattering data.  This would be
the analog of the SU(2) treatment of $\pi \pi$ scattering, where the
$\sigma$ alone can provide a reasonable description of the low energy
region.  The $\kappa$ cannot explain the data in the region of the $K_0^*(1430)$
scalar resonance.  In other words, we cannot explain the $K_0^*(1430)$
as the strange scalar of the usual linear sigma model treated with
K-matrix unitarization.  

In the case of the $\pi \eta$ channel there does not appear to be any
experimental phase shift data, so we compare with experimental
determinations of the $a_0(980)$ mass and width.  The lower physical
mass entry for the $a_0$ in Table \ref{summary_table} corresponds to
the bare mass of the renormalizable model.  It is somewhat too low but
not very far off.  This can be easily adjusted by using the
non-renormalizable potential.  The predicted width is somewhat too
large but qualitatively reasonable.  Clearly, the $a_0$ of the present
model is describing the low energy part of $\pi \eta$ scattering and
should correspond to the $a_0(980)$ rather than the $a_0(1450)$.

All in all, the three flavor linear sigma model with a general
(non-renormalizable) chiral invariant potential and regularized by the
simple K-matrix procedure can approximately describe the complicated
$\pi \pi$ scalar scattering amplitude as well as the low energy part
of the $K \pi$ scalar amplitude and the $a_0(980)$ $\pi \eta$ resonance.  The
$K_0^*(1430)$ and $a_0(1450)$ are ``outsiders'' in this picture and
would have to be put in by hand to realize the higher mass scalar
resonances in $\pi K$ and $\pi \eta$ scattering.  The picture is
qualitatively similar to that obtained in treatments using the
non-linear sigma model for $\pi \pi$ \cite{HSS1}, $\pi K$
\cite{BFSS1} and $\pi \eta$ \cite{BFS2} scattering.  The $a_0(980)$
and $f_0(980)$ seem to belong to the same multiplet as the
controversial light $\sigma$ and light $\kappa$.  Of course, it is
possible for particles with the same quantum numbers belonging to other
multiplets to mix with them.  

There are several straightforward, but lengthy to carefully implement,
ways to improve this treatment.  Modified kinetic terms, as mentioned
in section I, can be included to improved the fit to pseudoscalar
masses and decay constants.  Vector and axial vector mesons can be
added to introduce more of the low-lying physical resonances which are
expected to be important in the $\pi \pi$ and $\pi K$ channels.
Finally, inelastic channels can be included.

\section{Speculation on scalar meson's quark structure}

Up to this point we have reported the results of a straightforward and highly predictive treatment of the three flavor linear sigma model.  Our original reason for pursuing this investigation was to check the results obtained in our treatment of meson scattering in the non-linear sigma model which contained additional particles and channels.  That treatment used a different unitarization procedure in which crossing symmetry and unitarity were both approximately satisfied. (Actually in the study of direct channel scalar resonances, the crossed scalar exchanges are relatively small).  We already noted that the locations and widths of the {\it physical} scalar states obtained in the linear model were qualitatively similar to those obtained in the non-linear model.  Since the $\sigma$, $f_0$, $a_0$ and $\kappa$ all come out less than or about 1 GeV, and the scattering regions near the $a_0(1450)$ and the $K_0^*(1430)$ apparently must be described by fields other than those contained in the matrix $M$, the well known puzzle of the quark structure of these scalars comes to the surface. 

In this section we will make some speculative remarks on this controversial subject and introduce another toy model which may illuminate some of the issues.  The puzzle, of course, is why, if the scalars are ``$q \bar q$ states'', they are considerably lighter than the other p-wave states and why the isovector $a_0(980)$ is tied for being the heaviest, rather than the lightest, member of the multiplet. 

Actually there is a lot of ambiguity in stating what the quark structure of a physical hadron means.  Generally people think of the question in the context of a potential-type model wherein, for example, the $\rho$ meson is made of a ``constituent'' quark of mass about 300 MeV and a constituent anti-quark of the same mass.  The idea is that the fundamental ``current quarks of QCD'' (with masses about 10 MeV) interact strongly with each other and with gluons to make the relatively weakly interacting constituents whose combined masses roughly approximate the physical hadron masses.  Thus the quark structure really depends on the model used to treat the hadrons.  At the field theory level of ``current quarks'' there is always some probability for extra $q \bar q$ pairs or other structures to be present.  
In the $SU(3)_L \times SU(3)_R$ chiral effective Lagrangian
treatments, the quark substructure of the fields being used does not
enter the formulation in a unique way.  An infinite number of
different quark substructures will give rise to the same $SU(3)_L
\times SU(3)_R$ transformation properties for the mesons.  This is
apparent for the non-linear chiral model in which scalars are added to
the pseudoscalar meson Lagrangian as ``matter fields'' in the usual
manner \cite{CCWZ}.  Then it is known that only the SU(3) flavor
transformation properties of the scalars are relevant.  However we
found in our earlier study \cite{BFSS2} that the value of the scalar
mixing angle suggested indirectly that the light scalars do have an
important four quark component.  Considering the properties of the
heavier scalars $a_0(1450)$ and $K_0^*(1430)$ suggested \cite{BFS3}
that these states did not belong to a ``pure'' $q \bar q$ multiplet but to one which mixed with the lighter scalar multiplet. 

When it comes to the linear sigma model where the chiral transformations of the scalars are linked with those of the pseudoscalars in a natural way, there seems to be a feeling that the matrix $M$ should describe a $q \bar q$ field.  In fact, there are still an infinite number of quark substructures which transform in the same manner under $SU(3)_L \times SU(3)_R$.  It may be worthwhile to illustrate this for the specific cases of interest in the literature.  

The schematic structure for the matrix $M(x)$ realizing a $q \bar q$ composite in terms of quark fields $q_{aA}(x)$ can be written
\begin{equation}
M_a^{(1)b} = {\left( q_{bA} \right)}^\dagger \gamma_4 \frac{1 + \gamma_5}{2} q_{aA},
\label{M1}
\end{equation}
where $a$ and $A$ are respectively flavor and color indices.  Our convention for matrix notation is $M_a^{(1)b} \rightarrow M^{(1)}_{ab}$.  Then $M^{(1)}$ transforms under chiral $SU(3)_L \times SU(3)_R$ as
\begin{equation}
M^{(1)} \rightarrow U_L M^{(1)} U_R^\dagger
\end{equation}
where $U_L$ and $U_R$ are unitary, unimodular matrices associated with the transformations on the left handed ($q_L = \frac{1}{2}\left( 1 + \gamma_5 \right) q$) and right handed ($q_R = \frac{1}{2}\left( 1 - \gamma_5 \right) q$) quark projections.  For the discrete transformations charge congugation $C$ and parity $P$ one verifies 
\begin{equation}
C: \quad M^{(1)} \rightarrow  M^{(1)T}, \quad \quad P: \quad M^{(1)}({\bf x}) \rightarrow  M^{(1)\dagger}(-{\bf x}).
\end{equation}
Finally, the $U(1)_A$ transformation acts as $q_{aL} \rightarrow e^{i\nu} q_{aL}$, $q_{aR} \rightarrow e^{-i\nu} q_{aR}$ and results in:
\begin{equation}
M^{(1)} \rightarrow e^{2i\nu} M^{(1)}. 
\label{M1U1A}
\end{equation}

One interesting model \cite{Isgur} for explaining the scalar meson puzzle (at least insofar as the $a_0(980)$ and $f_0(980)$ states are concerned) is to postulate that the light scalars are ``molecules'' made out of two pseudoscalar mesons.  The chiral realization of this picture would result in the following schematic structure:
\begin{equation}
M_a^{(2)b} = \epsilon_{acd} \epsilon^{bef} {\left( M^{(1) \dagger} \right)}_e^c {\left( M^{(1) \dagger} \right)}_f^d.
\label{M2}
\end{equation}

One can verify that $M^{(2)}$ transforms exactly in the same way as $M^{(1)}$ under $SU(3)_L \times SU(3)_R$, $C$ and $P$.  Under $U(1)_A$ it transforms as 
\begin{equation}
M^{(2)} \rightarrow e^{-4i\nu} M^{(2)},
\end{equation}
which differs from Eq. (\ref{M1U1A}).  

Another interesting approach \cite{Jaffe} to explaining the light
scalar mesons was formulated by Jaffe in the framework of the MIT bag
model.  It was observed that the spin-spin (hyperfine) piece of the
one gluon exchange interaction between quarks gives an exceptionally
strong binding to an s-wave $qq\bar q \bar q$ scalar state.
Furthermore, this model naturally predicts an ``inverted'' mass
spectrum of the type summarized in Table \ref{summary_table}.  The scalar states of this
type may be formally written as bound states of a ``dual quark'' and
``dual antiquark''.  There are two possibilities if the dual antiquark
is required to belong to a $\bar 3$ representation of flavor SU(3).
In the first case it belongs to a $\bar 3$ of color and is a spin singlet.  This has the schematic chiral realization,
\begin{eqnarray}
L^{gE} = \epsilon^{gab} \epsilon^{EAB}q_{aA}^T C^{-1} \frac{1 + \gamma_5}{2} q_{bB}, \nonumber \\
R^{gE} = \epsilon^{gab} \epsilon^{EAB}q_{aA}^T C^{-1} \frac{1 - \gamma_5}{2} q_{bB}, 
\end{eqnarray}
where $C$ is the charge conjugation matrix of the Dirac theory.  A suitable form for the $M$ matrix is:
\begin{equation}
M_g^{(3)f} = {\left( L^{gA}\right)}^\dagger R^{fA}.
\end{equation}
$M^{(3)}$ can be seen to transform in the same way as $M^{(2)}$ under $SU(3)_L \times SU(3)_R$, $C$, $P$ and $U(1)_A$.  In the second case the dual antiquark belongs to a $6$ representation of color and has spin 1.  It has the corresponding schematic chiral realization:
\begin{eqnarray}
L_{\mu \nu,AB}^g = L_{\mu \nu,BA}^g = \epsilon^{gab} q^T_{aA} C^{-1} \sigma_{\mu \nu} \frac{1 + \gamma_5}{2} q_{bB}, \nonumber \\
R_{\mu \nu,AB}^g = R_{\mu \nu,BA}^g = \epsilon^{gab} q^T_{aA} C^{-1} \sigma_{\mu \nu} \frac{1 - \gamma_5}{2} q_{bB}, 
\end{eqnarray}
where $\sigma_{\mu \nu} = \frac{1}{2i} \left[ \gamma_\mu, \gamma_\nu \right] $.  This choice leads to an $M$ matrix
\begin{equation}
M_g^{(4) f} = {\left( L^{g}_{\mu \nu,AB}\right)}^\dagger R^{f}_{\mu \nu,AB},
\end{equation}
where the dagger operation includes a factor ${(-1)}^{\delta_{\mu 4} + \delta_{\nu 4}}$.  $M^{(4)}$ also transforms like $M^{(2)}$ and $M^{(3)}$ under all of $SU(3)_L \times SU(3)_R$, $C$, $P$ and $U(1)_A$.  The specific form favored by the MIT bag model calculation actually corresponds to a particular linear combination of $M^{(3)}$ and $M^{(4)}$.  Furthermore one can verify that $M^{(2)}$ in Eq. (\ref{M2}) is related by a Fierz transformation to a linear combination of $M^{(3)}$ and $M^{(4)}$.  Thus only two of $M^{(2)}$, $M^{(3)}$ and $M^{(4)}$ are linearly independent.  

What is the significance of these remarks for construction of the general effective chiral Lagrangian used in this paper [Eq. (\ref{LsMLag})]?  All that is required for $M$ is that it transform like $M^{(1)}$ under $SU(3)_L \times SU(3)_R$, $C$ and $P$ and that it carry a non-zero $U(1)_A$ ``charge'' which gets broken by the potential.  The specific $U(1)_A$ transformation property does differ between the two quark realization $M^{(1)}$ and the four quark realizations ($M^{(2)}$, $M^{(3)}$ and $M^{(4)}$) but this would just be absorbed, in the present work, by a different value for the parameter $V_4$.  Thus, if one knew nothing else about hadronic physics than the present toy Lagrangian, one would not be able to a priori easily discriminate among the possibilities $M^{(1)}-M^{(4)}$, or in fact any others, for the underlying quark substructure of the scalars (and pseudoscalars).  Nevertheless, one might glance at the obtained scalar masses in Table \ref{summary_table} and notice that there is an inverted physical mass spectrum.  One might then decide to make a judgement on the {\it constituent} quark substructure by fitting the scalar spectrum to an Okubo type mass formula \cite{Okubo}.  This was done recently, for example, in section II of \cite{BFSS2} and suggests that the scalars are behaving roughly as composites of four constituent quarks.  Roughly, this amounts to simply counting the number of strange constituent pieces in each state;  in the four quark picture both $f_0(980)$ and $a_0(980)$ have two.  The combined effects of spontaneous chiral symmetry breaking and unitarization (presumably taking radiative corrections into account) appears to split the constituent structures of the scalars from the pseudoscalars, regardless of which current quark structure (i.e. choice of $M$) we start with.  

However the true situation is likely to be more complicated.  The present model does not appear to accomodate the $a_0(1450)$ and $K_0^*(1430)$ scalars as states belonging to $M$.  These states would seem at first sight to be reasonable candidates for a nonet of ordinary $q \bar q$ scalars.  Still it is a little puzzling that $K_0^*(1430)$ is not heavier than $a_0(1450)$.  There are some other puzzles too but all can be qualitatively explained \cite{BFS3} if a $q \bar q$ scalar nonet mixes with a $qq\bar q \bar q$ scalar nonet.  If we want to realize such a scheme in the linear model framework it would be natural to introduce a Lagrangian with two different $M$ matrices.  Such a model seems to yield a variety of interesting dynamical possibilities which may lead to new insights and approximation schemes for low energy QCD.  Thus it may be worthwhile to give a brief discussion here.

Let us start with the field $M^{(1)}$ which we shall simply designate
$M$.  At the kinematical level it represents a current-type quark
antiquark operator.  This is modified for both the pseudoscalar and
scalar states by the (almost) spontaneous breakdown of chiral
symmetry.  For the scalars (which occur as poles in the physical
region) there is an additional modification due to the unitarization
required.  Of course, the choice of the free parameters gives an
``experimental'' input to this process. The resulting scalars seem to
be roughly consistent with a $qq\bar q  \bar q$ constituent-quark
structure, as just discussed.  Now consider adding a current type four
quark operator which may be any combination of $M^{(2)}$, $M^{(3)}$ or $M^{(4)}$ (we could not tell the difference in an effective Lagrangian framework) and denote it by $M^\prime$.  Allow $M^\prime$ to mix with $M$.  What happens?  

The Lagrangian which directly generalizes Eq. (\ref{LsMLag}) is written as
\begin{equation}
{\cal L} = - \frac{1}{2} {\rm Tr} \left( \partial_\mu M \partial_\mu M^\dagger \right) - \frac{1}{2} {\rm Tr} \left( \partial_\mu M^\prime \partial_\mu M^{\prime \dagger} \right) - V_0 \left( M, M^\prime \right) - V_{SB},
\label{mixingLsMLag}
\end{equation}
where $V_0(M,M^\prime) $ stands for a general polynomial made from $SU(3)_L \times SU(3)_R$ [but not $U(1)_A$] invariants formed out of $M$ and $M^\prime$.  Furthermore $V_{SB}$ is taken to be the same as Eq. (1.7) since it is ${\rm Tr} \left( M + M^\dagger \right)$ which ``mocks up'' the quark mass terms.  Other physical particles (including glueballs) could be added for more realism, but Eq. (\ref{mixingLsMLag}) is already quite complicated.

To get an indication of what kinds of questions might be answered, let us consider a very simplified approximation in which the quark mass effective term, $V_{SB}$ is absent and where $V_0$ is simply given by:
\begin{equation}
V_0 = -c_2 {\rm Tr} \left( M M^\dagger \right) + c_4 {\rm Tr} \left( M M^\dagger M M^\dagger \right) + d_2 {\rm Tr} \left( M^\prime M^{\prime \dagger} \right) + e {\rm Tr} \left( M M^{\prime \dagger} + M^\prime M^\dagger \right).
\label{mixingpot}
\end{equation}
Here $c_2$, $c_4$ and $d_2$ are positive real constants.  The $M$ matrix field is chosen to have a wrong sign mass term so that there will be spontaneous breakdown of chiral symmetry.  A pseudoscalar octet will thus be massless. On the other hand, the matrix field $M^\prime$ is being set up to have trivial dynamics except for its mixing term with $M$.  The mixing is controlled by the parameter $e$ and the $e$-term is the only one which violates $U(1)_A$ symmetry.  Its origin is presumably due to instanton effects at the fundamental QCD level.  (Other $U(1)_A$-violating terms which contribute to $\eta^\prime$ mass etc. are not being included for simplicity).  Using the notations $M=S + i \phi$ and $M^\prime = S^\prime + i \phi^\prime$ we may expect vacuum values
\begin{equation}
\left< S_a^b \right> = \alpha \delta_a^b, \quad \quad \left< S_a^{\prime b} \right> = \beta \delta_a^b.
\label{vevs}
\end{equation}
The minimization condition $\left< \frac{ \partial V_0}{\partial S_a^{\prime b}} \right> = 0$ leads to 
\begin{equation}
\beta = - \frac {e}{d_2} \alpha
\end{equation}
while $\left< \frac{ \partial V_0}{\partial S_a^{ b}} \right> = 0$ yields
\begin{equation}
\alpha^2 = \frac{1}{2c_4} \left( c_2 + \frac{e^2}{d_2} \right).
\label{Smin}
\end{equation}

In the absence of mixing the ``four-quark'' condensate $\beta$ vanishes while the usual two quark condensate $\alpha$ remains.  

The mass spectrum resulting from Eq. (\ref{mixingpot}) has two scalar octets and two pseudoscalar octets, each with an associated $SU(3)$ singlet.  Each octet has eight degenerate members since the quark mass terms have been turned off.  Let us focus on the I=1, positively charged particles for definiteness and define:
\begin{equation}
\pi^+ = \phi_1^2,\quad \pi^{\prime +} = \phi_1^{\prime 2}, \quad a^+ = S_1^2, \quad a^{\prime +} = S_1^{\prime 2}.
\end{equation}
Then the $2 \times 2$ squared mass matrix of $\pi $ and $\pi^\prime$ is:
\begin{equation}
2\left[ \begin{array}{c c}
\frac{e^2}{d_2} & e  \\
e & d_2 
\end{array}
\right].
\label{psmass}
\end{equation}
This has eigenstates
\begin{eqnarray}
\pi_p &= {\left( 1 + \frac{e^2}{d_2^2} \right)}^{- \frac{1}{2}} \left( \pi - \frac{e}{d_2} \pi^\prime \right), \nonumber \\
\pi_p^\prime &= {\left( 1 + \frac{e^2}{d_2^2} \right)}^{- \frac{1}{2}} \left( \frac{e}{d_2}\pi + \pi^\prime \right),
\end{eqnarray}
with masses
\begin{equation}
m^2 \left( \pi_p \right) = 0, \quad \quad m^2_{\rm \tiny BARE} \left( \pi^\prime_p \right) = \frac{2e^2}{d_2} + 2 d_2.
\end{equation}
We put the subscript ``BARE'' on $m^2 \left( \pi^\prime_p \right)$ to indicate that it may receive non-negligible corrections from K-matrix unitarization as in our detailed treatment of the $M$ only Lagrangian in the above.  A possible experimental candidate for such a particle is the $\pi(1300)$.  

Computing the axial vector current by Noether's theorem yields
\begin{eqnarray}
{\left( J_\mu^{\rm axial} \right)}_1^2 &=& F_\pi \partial_\mu \pi_p^+ + ... , \nonumber \\
F_\pi &=& 2\alpha \sqrt{ 1 + {\left( \frac{e}{d_2} \right)}^2},
\end{eqnarray}
where $\alpha$ is given in Eq. (\ref{Smin}).  

Notice that a term like $\partial _\mu \pi^{\prime +}_p$ does not appear in our semi-classical approximation.  

The $2 \times 2$ squared mass matrix of the scalars $a$ and $a^\prime$ is 
\begin{equation}
\left[
\begin{array}{c c}
4c_2 + \frac{6e^2}{d_2} & 2e \\
2e & 2d_2 
\end{array}
\right].
\end{equation}
The eigenstates are defined by 
\begin{equation}
\left( \begin{array}{c} a_p \\ a_p^\prime \end{array} \right) = 
\left[ \begin{array}{c c} {\rm cos}\omega & - {\rm sin} \omega \\
                         {\rm sin} \omega & {\rm cos} \omega
                         \end{array} \right] 
\left( \begin{array}{c} a \\ a^\prime \end{array} \right),
\end{equation}
with 
\begin{equation}
{\rm tan} 2\omega = \frac {e}{2d_2 - 4c_2 - \frac{6e^2}{d_2}}.
\label{tan_def}
\end{equation}
The corresponding masses are
\begin{equation}
m^2_{\rm \tiny BARE} \left( a_p, a_p^\prime \right) = 2c_2 + d_2 + \frac{3e^2}{d_2} \mp \frac{e}{2} \left( 3 {\rm sin} 2 \omega + {\rm csc} 2 \omega \right),
\end{equation}
where the upper (lower) sign stands for $a_p$, $\left( a_p^\prime \right)$.  

\begin{figure}
\centering
\epsfig{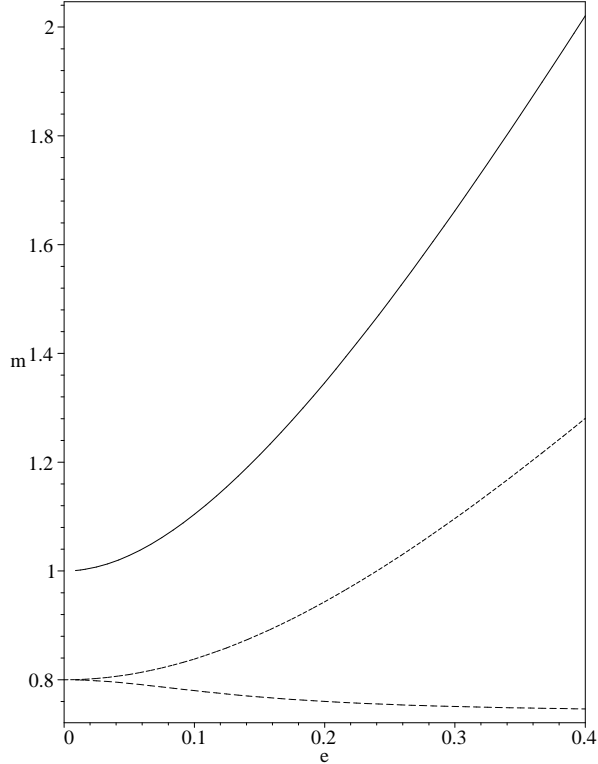}
\caption
{Plots of $m_{\rm \tiny BARE} \left( a_p \right)$ (solid), $m_{\rm \tiny BARE}
\left( a_p^\prime \right)$ (dashed) and $m_{\rm \tiny BARE}
\left( \pi_p^\prime \right)$ versus the mixing parameter $e$ for the
choice $c_2 = 0.25$ ${\rm GeV}^2$ and $d_2 = 0.32$ ${\rm GeV}^2$.  The
highest lying curve is mainly a ``$q\bar q$'' scalar, while the lowest
lying curve is mainly a ``$qq \bar q \bar q$'' scalar.  The excited
pseudoscalar curve is in the middle.}
\label{mixing_fig}
\end{figure}

It is interesting to examine the masses of the degenerate octets in a little more detail.  For orientation, first consider the case when the mixing parameter $e$ vanishes. The usual ``$q \bar q$'' pseudoscalars $\pi_p$ are zero mass Goldstone bosons in this approximation.  If $4c_2 > 2d_2$, $a_p$, the original scalar partner of $\pi_p$ lies higher than the degenerate ``$qq\bar q \bar q$'' scalar and pseudoscalar $a_p^\prime$ and $\pi_p^\prime$.   When the mixing is turned on, a four quark condensate develops and the mass ordering is
\begin{equation}
m_{\rm \tiny BARE} (a_p) > m_{\rm \tiny BARE} (\pi_p^\prime) > m_{\rm
\tiny BARE} (a_p^\prime) > m_{\rm \tiny BARE} (\pi_p) = 0.
\end{equation}
This is graphed, as a function of e, in Fig. \ref{mixing_fig} (with parameter choices $c_2 = 0.25$ ${\rm GeV}^2$, $d_2 = 0.32$ ${\rm GeV}^2$.  In such a scenario, the $qq \bar q \bar q$ scalar would be the next lightest after the $q \bar q$ Goldstone boson.  Each particle would be a mixture of $q \bar q$ and $qq \bar q \bar q$ to some extent.  For the given parameters the mixing angle remains small however because the denominator of Eq. (\ref{tan_def}) is always negative and increases in magnitude as $e^2$ increases.  Note especially, that due to the spontaneous breakdown of chiral symmetry, there is no guarantee that the lowest lying scalar is of $q \bar q$ type.  Also note that $\pi^\prime_p$ is expected to be more massive than $a_p^\prime$.  

On the other hand, if the QCD dynamics underlying the effective Lagrangian is such that $2d_2 > 4c_2$ we will get a mass ordering $m_{\rm \tiny BARE} (a_p^\prime) > m_{\rm \tiny BARE} (\pi^\prime) > m_{\rm \tiny BARE} (a_p)$ in which the four quark scalar appears heaviest.  However, in this case we will definitely get a large mixing as $e$ increases since the denominator of Eq. (\ref{tan_def}) starts out positive when $e=0$ and will go to zero as $e$ is increased.  Thus the next-to-lowest lying $a_p$ can be expected to have a large $qq \bar q \bar q$ admixture. 

All of these remarks pertain to the meson current-quark type operators
in the toy model.  The important effects of unitarization
(i.e. $m_{\rm \tiny BARE} \rightarrow m$) are likely, as in our earlier treatment, to favor an interpretation of the low lying physical scalars as being of four constituent quark type in either case.  

The main lesson from our preliminary treatment of a chiral model with mixing is perhaps that even though the $M$ fields carry ``chiral indices'' it is not easy to assign an unambiguous quark substructure.  On the other hand there is a great potentiality for learning more about non-perturbative QCD from further study of the light scalars.  Such features as scalar mixing (including the possibility of mixing with glueballs for the I=0 states), four quark condensates and excited pseudoscalars may eventually get correlated with each other and with the experimental data on the scattering of light pseudoscalars.

\acknowledgements
We are happy to thank Masayasu Harada and Francesco Sannino for many helpful discussions.  This work has been supported in part by the US DOE under contract DE-FG-02-85ER40231.

\appendix

\section{Coupling Constants and partial wave amplitudes}

For the scattering processes under consideration we will need the
four-point pseudoscalar contact interactions and the trilinear
scalar-pseudoscalar-pseudoscalar interactions.  In isotopic spin notation
the relevant pieces of the Lagrangian are respectively:
\begin{equation}
- {\cal L}^{\left( 4 \right)} = \frac{1}{16} g_\pi^{\left( 4 \right)} {\left( {\mbox{\boldmath
  ${\pi}$}} \cdot  {\mbox{\boldmath ${\pi}$}} \right)}^2 + \frac{1}{2}
  g_K^{\left( 4 \right)} {\bar K} K {\mbox{\boldmath
  ${\pi}$}} \cdot  {\mbox{\boldmath ${\pi}$}}   + \frac{1}{4}
  g_\eta^{\left( 4 \right)} {\eta} \eta {\mbox{\boldmath
  ${\pi}$}} \cdot  {\mbox{\boldmath ${\pi}$}} + ...
\end{equation}
and 
\begin{eqnarray}
-{\cal L}_{S\phi\phi} &=& \frac{g_{\kappa K \pi}}{\sqrt 2} \left(
 {\bar K} \mbox{\boldmath ${\tau}$} \cdot  {\mbox{\boldmath ${\pi}$}}
\kappa + h.c. \right) + \frac{g_{\sigma \pi \pi}}{2}
\sigma  \mbox{\boldmath ${\pi}$} \cdot 
{\mbox{\boldmath ${\pi}$}} + {g_{\sigma K  K}}\sigma  {\bar K} {K} +
 \frac{g_{\sigma^\prime \pi \pi}}{2} \sigma^\prime \mbox{\boldmath ${\pi}$} \cdot
 \mbox{\boldmath ${\pi}$} \\ \nonumber &+& {g_{\sigma^\prime K K}} \sigma^\prime {\bar K} {K} + \frac{g_{a_0 K K}}{\sqrt 2} {\bar K}
 \mbox{\boldmath ${\tau}$} \cdot {\bf a_0}  {K} +
g_{\kappa {K} \eta} \left(
{\bar \kappa}  K  {\eta} + h.c. \right) +
g_{\kappa {K} \eta '} \left(
{\bar \kappa} K  {\eta '} + h.c. \right) \\ \nonumber &+& g_{a_0 \pi\eta} {\bf a_0} \cdot
 \mbox{\boldmath ${\pi}$} \eta +
 g_{a_0 \pi\eta'} {\bf a_0} \cdot  \mbox{\boldmath ${\pi}$} \eta' + \frac{g_{\sigma \eta \eta}}{2} \sigma  \eta  \eta
+ g_{\sigma \eta \eta'} \sigma  \eta  \eta'
+ \frac{g_{\sigma \eta' \eta'}}{2} \sigma  \eta' \eta' \nonumber \\
&+&  \frac{g_{\sigma^\prime \eta \eta}}{2} \sigma^\prime  \eta  \eta
+  g_{\sigma^\prime \eta \eta'} \sigma^\prime  \eta  \eta'
+ \frac{g_{\sigma^\prime \eta' \eta'}}{2} \sigma^\prime \eta' \eta'.
\label{trilinear-interactions}
\end{eqnarray}
The trilinear couplings which do not involve three isoscalars are predicted
in terms of the masses.  These are given in \cite{SU1} and we
present them here for completeness:
\begin{eqnarray}
g_{\kappa K \pi}= \frac{1}{F_K} \left( m^2_{\rm \tiny BARE}(\kappa)   - m_\pi^2 \right), \quad
g_{\kappa K \eta} = \frac{1}{\sqrt{6}F_K}\left( {\rm cos} \theta_p +
2\sqrt{2}{\rm sin} \theta_p \right) \left( m_\eta^2 - m^2_{\rm \tiny BARE}(\kappa)  \right),
\\ \nonumber g_{\kappa K \eta^\prime}=\frac{1}{\sqrt{6}F_K}\left( 2\sqrt{2}{\rm cos} \theta_p -
{\rm sin} \theta_p \right) \left( m^2_{\rm \tiny BARE}(\kappa) - m_{\eta^\prime}^2 \right) \quad
g_{a_0K{K}} = \frac{1}{F_K} \left( m_{\rm \tiny BARE}^2(a_0) - m_K^2 \right), \\ \nonumber
g_{a_0\pi\eta} = \frac{\sqrt{2}}{F_\pi}a_p \left( m^2_{\rm \tiny BARE} (a_0) - m_\eta^2 \right),
\quad g_{a_0\pi\eta^\prime} = \frac{\sqrt{2}}{F_\pi}b_p \left( m^2_{\rm
\tiny BARE} (a_0)  -
m_{\eta^\prime}^2 \right), \\ \nonumber
g_{\sigma\pi\pi} = \frac{\sqrt{2}}{F_\pi}a_s \left( m^2_{\rm \tiny BARE} (\sigma) - m_\pi^2
\right), \quad  g_{\sigma^\prime \pi\pi} = \frac{\sqrt{2}}{F_\pi}b_s \left( m^2_{\rm \tiny BARE} (\sigma^\prime) - m_\pi^2
\right), \\ \nonumber g_{\sigma K {K}} = \frac{1}{\sqrt{6}F_K}\left( {\rm
cos}\theta_s + 2\sqrt{2} {\rm sin} \theta_s \right) \left( m_K^2 -
m^2_{\rm \tiny BARE} (\sigma) \right), \\ \nonumber
 g_{\sigma^\prime K { K}} = \frac{1}{\sqrt{6}F_K}\left( 2\sqrt{2}{\rm
cos}\theta_s - {\rm sin} \theta_s \right) \left(m^2_{\rm \tiny BARE} (\sigma^\prime) - m_K^2 \right).\\
\label{trilinear-couplings}
\nonumber
\end{eqnarray}
The trilinear coupling constants involving three isoscalars may depend on $V_4$.  For 
$\pi \eta$ elastic scattering we will also need:
\begin{equation}
g_{\sigma\eta\eta} = \frac{a_s}{\sqrt{2}}X - b_s Y, \quad  g_{\sigma^\prime\eta\eta} =
\frac{b_s}{\sqrt{2}} X + a_s Y, 
\end{equation}
where
\begin{eqnarray}
X = {\left(\frac{a_p}{\sqrt{2}} \right)}^2 \frac{2}{F_\pi} \left[ 2 a_s^2
m^2_{\rm \tiny BARE} (\sigma) + 2 {b_s^2}m^2_{\rm \tiny BARE}
(\sigma^\prime) - m_\pi^2 - a_p^2 m_\eta^2 - b_p^2m_{\eta^\prime}^2 -
12 (2F_K - F_\pi) V_4 \right]  \\ \nonumber 
 + {b_p}^2 \frac{2}{2F_K - F_\pi} \left[ - \sqrt{2} a_s
b_s \left( m^2_{\rm \tiny BARE} (\sigma) - m^2_{\rm \tiny BARE} (\sigma^\prime) \right) - 12 F_\pi V_4 \right]  +
\frac{48}{\sqrt{2}} a_p b_p V_4,  \nonumber
\end{eqnarray}
\begin{eqnarray}
Y= {\left(\frac{a_p}{\sqrt{2}} \right)}^2 \frac{2}{ F_\pi} \left[ -\sqrt{2}a_s
b_s \left( m^2_{\rm \tiny BARE} (\sigma) - m^2_{\rm \tiny BARE}
(\sigma^\prime) \right) - 24 F_\pi V_4 \right] \nonumber \\
 + b_p^2
\frac{2}{2F_K - F_\pi} \left[ b_s^2 m^2_{\rm \tiny BARE} (\sigma) +
a_s^2 m^2_{\rm \tiny BARE} (\sigma^\prime)  - b_p^2
m_\eta^2 - a_p^2 m_{\eta^\prime}^2 \right].  
\end{eqnarray}
In these equations we have used the convenient abbreviations 
\begin{equation}
a_p = \frac{{\rm cos} \theta_p
- \sqrt{2} {\rm sin} \theta_p} {\sqrt{3}}, \quad  b_p = \frac{ \sqrt{2}{\rm cos} \theta_p
+ {\rm sin} \theta_p}{\sqrt{3}},
\end{equation}
with analogous expressions for $a_s = {\rm cos} \psi$ and $b_s = {\rm sin} \psi$ in terms of $\theta_s$.
The contact coupling constants are then given by:
\begin{eqnarray}
g_\pi^{\left( 4 \right)} &=& \frac{4}{F_\pi^2} \left( a_s^2 m^2_{\rm \tiny BARE} (\sigma) +
b_s^2 m^2_{\rm \tiny BARE}(\sigma^\prime) - {m_\pi^2} \right), \\ \nonumber 
g_K^{\left( 4 \right)} &=& \frac{1}{F_\pi F_K} \left[ m^2_{\rm \tiny BARE}(\kappa) - m_K^2 -
m_\pi^2 + a_s^2 m^2_{\rm \tiny BARE} (\sigma) + b_s^2 m^2_{\rm \tiny
BARE}(\sigma^\prime) \right. \\ \nonumber
&-& \left. \sqrt{2} a_s b_s
\left(m^2_{\rm \tiny BARE} (\sigma) - m^2_{\rm \tiny BARE}(\sigma^\prime) \right) \right], \\ \nonumber 
g_\eta^{\left( 4 \right)}&=& \frac{2}{F_\pi} \left
[ \frac{a_s}{\sqrt{2}}g_{\sigma\eta\eta}+
\frac{b_s}{\sqrt{2}}g_{\sigma^\prime\eta\eta}+\frac{2}{F_\pi} a^2_p
\left(m^2_{\rm \tiny BARE}(a_0)
- {m_\eta}^2 \right) \right]. \\ 
\label{contactcouplings}
\end{eqnarray}

Finally, the tree-level partial wave amplitudes for $\pi K$ and $\pi
\eta$ scattering are:
\begin{eqnarray}
T_{0tree}^{1/2} &=& \rho (s) \left[ -2 g_K^{(4)} + g_{\kappa K \pi}^2 \left[ -\frac{1}{4q^2} {\rm ln} \left(
\frac{B_K + 1}{B_K - 1} \right) + \frac{3}{m^2_{\rm \tiny
BARE}(\kappa) - s} \right] \right. \\ \nonumber
&+&
\left. \frac{1}{2q^2}g_{\sigma\pi\pi}g_{\sigma KK} {\rm ln} \left( \frac
{m^2_{\rm \tiny BARE} (\sigma) + 4q^2}{m^2_{\rm \tiny BARE} (\sigma)}
\right) + \frac{1}{2q^2}g_{\sigma^\prime\pi\pi}g_{\sigma^\prime KK} {\rm ln} \left( \frac
{m^2_{\rm \tiny BARE} (\sigma^\prime) + 4q^2}{m^2_{\rm \tiny BARE} (\sigma^\prime)}
\right)\right]
\end{eqnarray}
and
\begin{eqnarray}
T_{0tree}^{1} &=& \rho (s) \left[ -2 g_\eta^{(4)} + g_{a_0 \pi \eta }^2 \left[ \frac{1}{2q^2} {\rm ln} \left(
\frac{B_\eta + 1}{B_\eta - 1} \right) + \frac{2}{m^2_{\rm \tiny
BARE}(a_0) - s} \right] \right. \\  \nonumber
&+& \left.
\frac{1}{2q^2}g_{\sigma\pi\pi}g_{\sigma \eta \eta} {\rm ln} \left( 1 +
\frac{4q^2}{m^2_{\rm \tiny BARE} (\sigma)}
\right) + \frac{1}{2q^2}g_{\sigma^\prime\pi\pi}g_{\sigma^\prime \eta \eta} {\rm ln} \left(1+ \frac{4q^2}{m^2_{\rm \tiny BARE} (\sigma^\prime)}
\right) \right],
\end{eqnarray}
where $q(s)$ and $\rho(s)$ for each case are given by Eqs. (\ref{cofm})
and  (\ref{kinematicalfactor}) respectively.  Furthermore
\begin{equation}
B_K = \frac{1}{2q^2} \left[ m^2_{\rm \tiny BARE}(\kappa) - m_\pi^2 -
m_K^2 + 2 \sqrt{ \left( m_\pi^2 + q^2 \right) \left( m_K^2 + q^2
\right) } \right]
\end{equation}
and 
\begin{equation}
B_\eta = \frac{1}{2q^2} \left[ m^2_{\rm \tiny BARE}(a_0) - m_\pi^2 -
m_\eta^2 + 2 \sqrt{ \left( m_\pi^2 + q^2 \right) \left( m_\eta^2 + q^2
\right) } \right].
\end{equation}


\begin{thebibliography}{10}

\bibitem{vanBev} E. van Beveren, T.A. Rijken, K. Metzger,
C. Dullemond, G. Rupp and J.E. Ribeiro, Z. Phys. {\bf C30}, 615
(1986). E. van Beveren and G. Rupp, hep-ph/9806246, 248.  See also
J.J. de Swart, P.M.M. Maessen and T.A. Rijken, U.S./Japan Seminar on the 
YN Interaction, Maui, 1993 [Nijmegen report THEF-NYM 9403].

\bibitem{MP} 
D. Morgan and M. Pennington, Phys. Rev. {\bf D48},  1185  (1993).

\bibitem{BMPV} A.A. Bolokhov, A.N. Manashov, M.V. Polyakov and
V.V. Vereshagin, Phys. Rev. {\bf D48}, 3090 (1993).  See also
V.A. Andrianov and A.N. Manashov, Mod. Phys. Lett. {\bf A8}, 2199
(1993).  Extension of this string-like approach to the $\pi K$ case
has been made in V.V. Vereshagin, Phys. Rev. {\bf D55}, 5349 (1997)
and very recently in A.V. Vereshagin and V.V. Vereshagin, {\it{ibid.}} {\bf
59}, 016002 (1999) which is consistent with a light $\kappa$ state.

\bibitem{AS} N.N. Achasov and G.N. Shestakov, Phys. Rev. {\bf
D49}, 5779 (1994).

\bibitem{Kam94}{R. Kam\'inski}, {L. Le\'sniak} and J. P. Maillet,
Phys. Rev. {\bf D50}, 3145 (1994).

\bibitem{SS}F.~Sannino and J.~Schechter, Phys. Rev.  {\bf D52},  96  (1995).

\bibitem{T}{N.A.~T\"ornqvist}, Z. Phys. 
{\bf C68}, 647 (1995) and references therein.  In addition see
{N.A.~T\"ornqvist} and M. Roos, Phys. Rev. Lett. {\bf 76}, 1575
(1996), N.A. T\"ornqvist, hep-ph/9711483 and Phys. Lett. {\bf B426} 105 (1998).

\bibitem{DS} R. Delbourgo and M.D. Scadron, Mod. Phys. Lett. {\bf
A10}, 251 (1995).  See also D. Atkinson, M. Harada and A.I. Sanda,
Phys.~Rev. {\bf D46}, 3884 (1992).

\bibitem{JPHS}
{G.~Janssen, B.C.~Pearce, K.~Holinde and J.~Speth}, Phys. Rev. {\bf  
D52},  2690  (1995).

\bibitem{Sv} M. Svec, Phys. Rev. {\bf D53}, 2343 (1996). 

\bibitem{Ishida} S. Ishida, M.Y. Ishida, H. Takahashi, T. Ishida,
K. Takamatsu and T Tsuru, Prog. Theor. Phys. {\bf 95}, 745 (1996), 
S.~Ishida, M.~Ishida, T.~Ishida, K.~Takamatsu and T.~Tsuru,
Prog. Theor. Phys. {\bf 98}, 621 (1997). See also M. Ishida and S. Ishida,
Talk given at 7th International Conference on Hadron Spectroscopy (Hadron
97), Upton, NY, 25-30 Aug. 1997, hep-ph/9712231. 

\bibitem{HSS1}M. Harada, F. Sannino and J. Schechter, Phys. Rev. {\bf D54},  
1991 (1996).

\bibitem{HSS2}M. Harada, F. Sannino and J. Schechter, Phys. Rev. Lett. {\bf 78}, 1603 (1997).

\bibitem{BFSS1}D. Black, A.H. Fariborz, F. Sannino and J. Schechter,
Phys. Rev. {\bf D58}, 054012 (1998). 

\bibitem{BFSS2}D. Black, A.H. Fariborz, F. Sannino and J. Schechter, Phys. Rev. {\bf D59}, 074026 (1999).

\bibitem{OOP} J.A. Oller, E. Oset and J.R. Pelaez, Phys. Rev. Lett.
{\bf 80}, 3452 (1998). See also K. Igi and K. Hikasa, Phys. Rev. {\bf
D59}, 034005 (1999).

\bibitem{AnSa}A.V. Anisovich and A.V. Sarantsev, Phys. Lett. {\bf B413},
137 (1997).

\bibitem{EFSS}V. Elias, A.H. Fariborz, Fang Shi and T.G. Steele,
Nucl. Phys. {\bf A633}, 279 (1998).

\bibitem{Dm} V. Dmitrasinovi\'c, Phys. Rev. {\bf C53}, 1383 (1996).

\bibitem{MO}P. Minkowski and W. Ochs, Eur. Phys. J. C {\bf 9}, 283 (1999).

\bibitem{GN}S. Godfrey and J. Napolitano, hep-ph/9811410.

\bibitem{BG}L. Burakovsky and T. Goldman, Phys. Rev. {\bf D57}
2879 (1998)

\bibitem{BFS1}D. Black, A. H. Fariborz and J. Schechter, Phys. Rev
{\bf D59}, 074026 (1999).


\bibitem{BFS2}D. Black, A. H. Fariborz and J. Schechter,
Phys. Rev. {\bf D61} 074030 (2000).  See also V. Bernard, N. Kaiser and U-G. Meissner, Phys. Rev. D {\bf 44}, 3698 (1991) and A. H. Fariborz and J. Schechter, Phys. Rev. {\bf D60}, 034002 (1999). 

\bibitem{BFS3}D. Black, A. H. Fariborz and J. Schechter, Phys. Rev. {\bf D61} 074001 (2000).

\bibitem{Shakin}L. Celenza, S-f Gao, B. Huang and C.M. Shakin, Phys. Rev. C {\rm 61}, 035201 (2000).

\bibitem{proc}An up to date summary of the status of the light scalar mesons will soon be available as the Proceedings of the ``Sigma workshop'', Kyoto, Japan, June 2000.  

\bibitem{CPT}J. Gasser and H. Leutwyler, Ann. Phy. (NY) {\bf 158}, 142, (1984); Nucl. Phy. {\bf B250}, 465, (1985).  A more recent review is given by U-G. Meissner, Rep. Prog. Phys. {\bf 56}, 903 (1993).

\bibitem{LargeN} E. Witten, Nucl. Phys. {\bf B160}, 57 (1979). See also
S. Coleman, {\it Aspects of Symmetry}, Cambridge University Press
(1985). The original suggestion is given in G. 't Hooft, Nucl. Phys. {\bf
B72}, 461 (1974).

\bibitem{Other}See the references \cite{vanBev}-\cite{proc} above. 

\bibitem{GL}M. Gell-Mann and M. L\'evy, Nuovo Cimento {\bf 16}, 705 (1960).

\bibitem{Weinberg}J. Cronin, Phys. Rev. {\bf 161}, 1483 (1967); S. Weinberg, Phys. Rev. Lett {\bf 18}, 188 (1967).

\bibitem{Hatsuda}T. Hatsuda, T. Kunihiro and H. Shimizu, Phys. Rev. Lett. {\bf 82}, 2840 (1999); S. Chiku and T. Hatsuda, Phys. Rev. D {\bf 58}, 076001 (1998).

\bibitem{CH}See also L.H. Chan and R. W. Haymaker, Phys. Rev. {\bf 07}, 402 (1973); {\bf 10}, 4170 (1974).

\bibitem{Chung}See for example, S. U. Chung {\it et al}, Ann. Physik {\bf 4} 404 (1995).  

\bibitem{Levy}M. L\'evy, Nuovo Cimento {\bf 52A}, 23 (1967).  See S. Gasiorowicz and D. A. Geffen, Rev. Mod. Phys. {\bf 41}, 531 (1969) for a review which contains a large bibliography.  



\bibitem{SU1}J. Schechter and Y. Ueda, Phys. Rev. {\bf D3}, 2874, 1971; Erratum D {\bf 8} 987 (1973).  See also J. Schechter and Y. Ueda, Phys. Rev. {\bf D3}, 168, (1971).

\bibitem{GJJS}H. Gomm, P. Jain, R. Johnson and J. Schechter, Phys. Rev. {\bf D33}, 801 (1986).

\bibitem{MS}V. Mirelli and J. Schechter, Phys. Rev. D {\bf 15}, 1361 (1977).

\bibitem{SU2}J. Schechter and Y. Ueda, Phys. Rev. D {\bf 4}, 733 (1971).

\bibitem{HS}W. Hudnall and J. Schechter, Phys. Rev. {\bf D9}, 2111,
1974.  See footnote in section I of the present paper for corrections of typographical errors.

\bibitem{W}S. Weinberg, Phys. Rev. Lett {\bf 17}, 616 (1966).

\bibitem{CCWZ}C. Callan, S. Coleman, J. Wess and B. Zumino, Phys. Rev. {\bf 177}, 2247 (1969).

\bibitem{pipidata}E.A. Alekseeva {\it et al}., Sov. Phys. JETP {\bf
55}, 591 (1982), G. Grayer {\it et al}., Nucl. Phys. {\bf B75}, 189 (1974). 

\bibitem{Aston}D. Aston {\it et al}., Nucl. Phys. {\bf B296}, 493 (1988).

\bibitem{PDG}Particle Data Group, Eur. Phys. J. C{\bf 15}, 1 (2000).

\bibitem{Isgur}N. Isgur and J. Weinstein, Phys. Rev. Lett {\bf 48}, 659
(1982); Phys. Rev. {\bf D27} 588 (1983).  See also F.E. Close, N. Isgur
and S. Kumano, Nucl. Phys. {\bf{B389}}, 513 (1993).

\bibitem{Jaffe}R.L. Jaffe, Phys. Rev. {\bf D15}, 267 (1977); See also N. Achasov and V. N. Ivanchenko,  Nucl. Phy. B{\bf315}, 465 (1989);  M.N. Achasov et al, Phys. Lett. {\bf{B 440}}, 442 (1998) 

\bibitem{Okubo}S. Okubo, Phys. Lett. {\bf 5}, 165 (1963).

\end{thebibliography}
\end{document}